\documentclass{aa}
\usepackage[fleqn]{amsmath}
\usepackage{graphicx,natbib,txfonts,color,upgreek}
\bibpunct{(}{)}{;}{a}{}{,}

\begin{document}
\title{Kalman-filter control schemes for fringe tracking}
\subtitle{Development and application to VLTI/GRAVITY}
\author{J. Menu\inst{1}\fnmsep\inst{2}\fnmsep\inst{3}\fnmsep\thanks{PhD fellow of the Research Foundation -- Flanders (FWO)} \and G. Perrin\inst{1}\fnmsep\inst{3} \and E.
Choquet\inst{1}\fnmsep\inst{3} \and S. Lacour\inst{1}\fnmsep\inst{3}}

\institute{ 
LESIA, Observatoire de Paris, CNRS, UPMC, Universit\'e Paris Diderot, Paris Sciences et Lettres, 5 place Jules Janssen, 92195 Meudon, France \and
Instituut voor Sterrenkunde, KU Leuven, Celestijnenlaan 200D, 3001 Leuven, Belgium \\ \email{jonathan.menu@ster.kuleuven.be} 
\and 
Groupement d'Int\'er\^et Scientifique PHASE (Partenariat Haute r\'esolution Angulaire Sol Espace) between ONERA, Observatoire de Paris, CNRS and Universit\'e Paris Diderot, France }
\date{Received / Accepted}

\authorrunning{J.~Menu et al.}
\titlerunning{Kalman-filter control schemes for fringe tracking}

\abstract
{The implementation of fringe tracking for optical interferometers is inevitable when optimal exploitation of the instrumental capacities is desired. Fringe tracking allows continuous fringe observation, considerably increasing the sensitivity of the interferometric system. In addition to the correction of atmospheric path-length differences, a decent control algorithm should correct for disturbances introduced by instrumental vibrations, and deal with other errors propagating in the optical trains. }
{In an effort to improve upon existing fringe-tracking control, especially with respect to vibrations, we attempt to construct control schemes based on Kalman filters. Kalman filtering is an optimal data processing algorithm for tracking and correcting a system on which observations are performed. As a direct application, control schemes are designed for GRAVITY, a future four-telescope near-infrared beam combiner for the Very Large Telescope Interferometer (VLTI).}
{We base our study on recent work in adaptive-optics control. The technique is to describe perturbations of fringe phases in terms of an \emph{a priori} model. The model allows us to optimize the tracking of fringes, in that it is adapted to the prevailing perturbations. Since the model is of a parametric nature, a parameter identification needs to be included. Different possibilities exist to generalize to the four-telescope fringe tracking that is useful for GRAVITY.}
{On the basis of a two-telescope Kalman-filtering control algorithm, a set of two properly working control algorithms for four-telescope fringe tracking is constructed. The control schemes are designed to take into account flux problems and low-signal baselines. First simulations of the fringe-tracking process indicate that the defined schemes meet the requirements for GRAVITY and allow us to distinguish in performance. In a future paper, we will compare the performances of classical fringe tracking to our Kalman-filter control.}
{Kalman-filter based control schemes will likely become the next standard for fringe tracking, providing the possibility to maximize the tracking performance. The results of the present study are currently being incorporated into the final design of GRAVITY.}


\maketitle


\section{Introduction}
Ever since the early days of optical interferometry, the motion of fringes due to atmospheric turbulence has been a major limitation to this observation technique. One possible way to avoid the blurring of fringes on the detector is to use short fringe exposures, since short integration times temporally ``freeze'' atmospheric motion. In combination with the low optical throughput of interferometric systems, however, the sensitivity associated with short integrations is low, even on the largest-telescope systems. 

Tests with prototype interferometers (\citealt{1980ApOpt..19.1519S}; for a further overview, see \citealt{2003RPPh...66..789M}) demonstrated the use of the control technique \emph{fringe tracking} to increase the sensitivity of interferometric observations. Correcting for the atmospherically introduced optical path difference (OPD) between different interferometer arms indeed allows the continuous observation of the fringes. In a fringe tracker, a phase sensor estimates the OPD corresponding to the current fringe exposure. On the basis of this value, a command is calculated by the OPD controller, which is then sent to an OPD-correcting actuator system. The result is that fringes in the scientific beam combiner are stabilized to atmospheric motion, and can be integrated much longer than the typical atmospheric coherence time. In addition, the synchronous tracking of the fringes on a bright source allows the interferometric observation of much fainter sources.

Apart from turbulence, an important limitation of current real-time wavefront correction or tracking is instrumental vibrations. Initial on-sky tests of NAOS, the first adaptive-optics (AO) system on the Very Large Telescope (VLT), attributed a loss in Strehl ratio of up to $15\%$ to mechanical vibrations \citep{2003SPIE.4839..140R}. In the case of the VLT Interferometer (VLTI), longitudinal vibrations on the level of the baselines cause fringes to move onto the detector. Power spectra of sequences of OPD values estimated during a run on the VLTI/PRIMA instrument \citep{2008NewAR..52..199D} indicate the presence of a discrete number of vibration components during fringe observations \citep{2009A&A...507.1739S}. In an effort to reduce the vibration level of the VLTI system, a campaign to identify the contribution of different VLT subsystems has been initiated \citep{2008SPIE.7013E..11H,2010SPIE.7734E.101P}. The efforts to isolate noisy instruments has led to significant improvements.

Active correction techniques can also be applied to reduce vibration contribution. On the one hand, a vibration-sensing accelerometer system operates on the first three mirrors of the four 8.2-m unit telescopes (UTs) and effectively corrects vibrations in the range 10-30\,Hz \citep{2008SPIE.7013E..11H}. On the other hand, the control algorithm of the OPD controller can be extended with additional vibration blocks that correct for a defined set of vibration components. The Vibration TracKing algorithm (VTK; \citealt{2008SPIE.7013E..15D}) is such an example that operates as a narrow-band filter. For the Keck Interferometer, a similar narrow-band control system has been designed \citep{2010PASP..122..795C}.

In recent years, there has been an effort in AO to design new control schemes. Of special importance are the techniques based on Kalman filtering \citep{1960TASME..82.35K}, which is a technique designed for general data processing. It is used to track and correct observed systems in a statistically optimal way, in that it minimizes the norm of the residual signal (for linear systems with Gaussian noise statistics). \citet{2004JOSAA..21.1261L} designed an AO control scheme based on Kalman filtering. One of the important characteristics of this scheme is that it can be extended \citep{2006PhDT.........2P} to coherently treat turbulence and vibrations, rather than add extra vibration control blocks to an unsuitable classical controller. As usual for Kalman filters, it is based on an \emph{a priori} model for the controlled system, which here is referred to as turbulence and vibration perturbations.

The purpose of this work is the design of Kalman-filtering control schemes for fringe tracking, starting from the work that has been done for AO control. In particular, we consider the case of four-telescope interferometry, and develop control schemes that will be applicable to the fringe-tracking subsystem of GRAVITY \citep{2011Msngr.143...16E,2005AN....326..561E}, which is a four-telescope beam combiner to be installed on the VLTI. The main purpose of this second-generation instrument is high-precision narrow-angle astrometry and phase-referenced interferometric imaging in the near-infrared K band. The main scientific target of GRAVITY is Sgr A$^\ast$, which is currently the most well-studied and likely supermassive black hole, located at the Galactic center. On-sky operation is expected in 2014.

Conceptually, the idea of using Kalman filtering to correct for atmospheric fringe motion dates back to \citet{1990SPIE.1237..172R}, but has never been implemented under that form. A simple Kalman-filter based strategy for single-baseline fringe tracking was effectively used at the Palomar Testbed Interferometer (PTI, \citealt{1999ApJ...510..505C}), which to our knowledge, is the only example of on-sky fringe tracking using Kalman-filter techniques. \citet{2010SPIE.7734E..82L} demonstrated the operation of Kalman filtering for vibration correction on a laboratory prototype for single-baseline space-based nulling interferometry.

The use of Kalman filtering for (ground-based) fringe tracking will have several advantages compared to classical control strategies (e.g.~integrator control):
\begin{itemize}
 \item Kalman filtering allows us to control turbulence and longitudinal vibrations in a coherent and equivalent way: the contribution of different perturbation components is decomposed.
 \item All information on disturbances is used via an \emph{a priori} model. This makes Kalman filtering a statistically optimal control method when considering linear systems with Gaussian white noise statistics.
 \item The use of Kalman filters may solve issues with existing active vibration control at VLTI (e.g.~artifacts in frequency spectra of residual VTK-corrected data; see \citealt{2008SPIE.7013E..11H}).
 \item The predictive nature of the filter allows us to deal with the loss of observables due to measurement noise.
\end{itemize}
In the case of the latter point, fringe-tracking control schemes need to be robust against flux losses owing to imperfect beam combination. Failures in tip-tilt correction, for instance, can lead to the short-time absence of fringe signal. The prediction capability of the Kalman filter controller is leveraged for this purpose.

The outline of this work is the following. In Sect.~\ref{sect:Kalman2}, we present the AO Kalman-filter control scheme, translated into a form applied to two-telescope fringe tracking. The Kalman filter is based on a parametric model and we present a technique for parameter identification in Sect.~\ref{sect:identification}. Since an aim of this work is to develop fringe-tracker control schemes applicable to GRAVITY, we extend the control in Sect.~\ref{sect:Kalman2} to four-telescope systems (Sect.~\ref{sect:Kalman4}). After considering some specific improvements in Sect.~\ref{sect:viszero}, we introduce the complete control schemes in Sect.~\ref{sect:full}. Before concluding this work, the results of the first simulations of four-telescope Kalman-filter fringe tracking for GRAVITY are presented (Sect.~\ref{sect:simulations}).

\section{Basic case: two-telescope Kalman-filter control} \label{sect:Kalman2}
The input information provided for fringe-tracking control are the OPD values estimated by the phase sensor. The aim of the OPD controller is to give an appropriate output command to the piston actuators (piezo actuators and delay lines). This section introduces the control law that is appropriate for a two-telescope interferometer. The approach used is based on a formalism developed for AO modeling, hence we can directly apply it here to fringe tracking. For a more complete introduction to this model, we refer to \citet{2004JOSAA..21.1261L} and \citet{2006PhDT.........2P}, which are the papers on which the current section is based. The notation we use is (mainly) that of \citet{2010JOSAA..27.122M}.

\subsection{Modeling the fringe-tracking control}\label{sect:modeling}
We have already mentioned before that Kalman filters need a model for the controlled system. We successively introduce the model for the fringe tracking and the evolution of the disturbances (turbulence and vibrations).

\subsubsection{Description of the fringe-tracking process}
Turbulence and longitudinal vibrations add up to introduce OPD perturbations. The OPD in a two-telescope interferometer can be related to a phase value $\varphi$ as
\begin{equation}\varphi\equiv\varphi^\mathrm{tur}+
\sum_i\varphi^{\mathrm{vib}\,i}=\frac{2\pi}\lambda\mathrm{OPD}.
\end{equation}
In reality, the quantity $y$ observed by a fringe tracker is not exactly $\varphi$, but contains two additional components. The first of these is an additive measurement noise $w$. Secondly, a fringe tracker in operation will apply correction phases, or \emph{commands}, $u$ using the piston actuators. We can therefore express the measured value of the fringe tracker as
\begin{equation}
y=\varphi-u+w.\label{eq:simpley}
\end{equation}
In what follows, we refer to $y$ as the residual OPD. For simplicity, we ignore conversion factors, e.g.~$2\pi/\lambda$, and assume that all quantities are expressed in the same units (say, microns).\footnote{More generally, one can write
$y=D\,\big(\varphi-N\,u\big)+w$, with conversion factors $D$ and $N$.
}

\begin{figure}
\centering
\includegraphics[width=0.49\textwidth,viewport=140 650 480 730,clip]{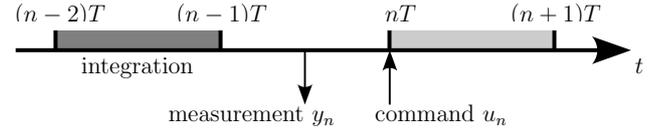}
\caption{Time diagram of the fringe tracking process.}\label{fig:scheme}
\end{figure}

To get into the problem of discretization, we denote by $T$ the elementary integration time of the fringe tracker. The discretized variables are then defined in the following way:
\begin{itemize}
\item $\varphi^i_n$ is defined as the average phase of the disturbance component $i$ during the time interval $[(n-1)T, nT]$:
\begin{equation}
\varphi^i_n\equiv\frac1T\int_{(n-1)T}^{nT}\mathrm
dt\,\varphi^i(t).
\end{equation} The sum of all these phases is referred to as
$\varphi_n$.
\item The command is a fixed value during one integration (assuming that the response time of the actuators is much shorter than $T$). We therefore define $u_n$ as the command applied at time step $n$, i.e.~applied during the time interval $[nT, (n+1)T]$.
\item The final parameter, the noisy component $w_n$ at time step $n$, could in principle be defined in a similar way as $\varphi^i_n$. In this model, however, $w$ will be considered as simple zero-mean white Gaussian noise that adds to $y$. The standard deviation is fixed to $\sigma_w$, which makes the values $w_n$ independent of other variables. 
\end{itemize}
The assumption that the data processing after integration amounts to one time step $T$ leads to the following discretized version of Eq.~(\ref{eq:simpley}):
\begin{equation}
y_n=\varphi_{n-1}-u_{n-2}+w_n.\label{eq:yn}
\end{equation}
The quantity $y_n$ is thus the measurement that is available at time step $n$. We include a schematic representation of the fringe measurement model in Fig.~\ref{fig:scheme}.

A final note to this model for the fringe-tracking process concerns the notion of a fixed measurement error $\sigma_w$. Further on in this work (Sect.~\ref{sect:viszero}), we introduce an adaptive weighting according to the (variable) signal-to-noise ratio (S/N) conditions. The current assumption of a fixed $\sigma_w$ is only for clarity reasons and not for the actual implementation.

\subsubsection{Description of the OPD evolution}\label{sect:descriptionOPDev} 
The formalism of Kalman filtering that is used to estimate the commands, requires an \emph{a priori} model not only for the measurement process, but also for the evolution of the observed system, that is, of the turbulence and the longitudinal vibrations. In essence, this requires the description of new states of the system in terms of previous realizations.

The OPD spectra obtained at the VLTI typically contain a number of peaks (e.g~\citealt{2009A&A...507.1739S}), which are associated with vibration modes excited by a variety of processes. In the following, we can assume each mode to be independent. The simplest way is then to describe the dynamics of such a vibration mode $i$ is to consider it as a discrete damped oscillator, excited by a Gaussian white noise $v$. It can be shown \citep{2006PhDT.........2P} that the one-dimensional equation of motion for a vibration mode $i$ of natural frequency $f_0^i$ and damping coefficient $k^i$ can be discretized into the recursive equation
\begin{equation}
\varphi^{\mathrm{vib}\,i}_{n+1}=a^{\mathrm{vib}\,i}_1\varphi^{\mathrm{vib}\,i}_n+
a^{\mathrm{vib}\,i}_2\varphi^{\mathrm{vib}\,i}_{n-1}+v^{\mathrm{vib}\,i}_n, \label{eq:phidiscrevo}
\end{equation}
where
\begin{eqnarray}
a^{\mathrm{vib}\,i}_1&=&2\,e^{-2\pi\,k^if_0^iT}\,\cos\Big(2\pi\,f^i_0T\sqrt{1-{k^i}^2}\Big),\\
a^{\mathrm{vib}\,i}_2&=&-e^{-4\pi\,k^if^i_0T}.\label{eq:aparameters}
\end{eqnarray}
Eq.~(\ref{eq:phidiscrevo}) has a form that is better known as an autoregressive model of order two, AR(2). We note that the conversion from continuous to discrete involves a Taylor approximation.

The modeling of longitudinal vibrations as randomly-excited damped oscillators leads in a natural way to an AR(2) description. One may ask whether the contribution of turbulence can also be described as an AR$(p)$ process. \citet{2004JOSAA..21.1261L} use an AR(1) turbulence model, based on temporal evolution characteristics of the phenomenon and basic parameters such as wind speed and the telescope diameter. A slightly more advanced first-order temporal model is discussed in \citet{2007JOSAA..24..2850L}.

When taking a damping coefficient $k>1$, the previous AR(2) vibration model can also be used to describe non-resonating signals, as in the case of turbulence.\footnote{The complex identities $\sqrt{1-k^2}=\mathrm j\sqrt{k^2-1}$ and $\cos(\mathrm j\theta)=\cosh\theta$ are used to compute $a_1$ in Eq.~(\ref{eq:aparameters}).} This property led \citet{2010JOSAA..27.122M} to use an AR(2) model for the turbulence description. We follow their argumentation and write
\begin{equation}
\varphi^{\mathrm{tur}}_{n+1}=a^\mathrm{tur}_1\varphi^{\mathrm{tur}}_n+a^\mathrm{tur}_2\varphi^{\mathrm{tur}}_{n-1}+v^\mathrm{tur}_n.\label{eq:phidiscrtur}
\end{equation}

\subsection{State-space description of the system control}\label{sect:statespace}
The state-space representation is a general framework to describe problems involving an evolving system on which measurements are performed (for a general introduction, see \citealt{2004Stsp.book.....D}). Under the assumption of linearity and time invariance (LTI state-space models), the model has the form of two recursive (constant) matrix equations: one describes the evolving system and the other the observational process on that system.

At least qualitatively, we have good indications that the considered problem can be described as an LTI problem. On the one hand, the states of turbulence and vibrations can be considered as realizations of a time-invariant system (for reasonable timescales). On the other hand, the OPD measurements are clearly a form of observation of that system. Using standard state-space notation, we can cast Eqs.~(\ref{eq:yn}), (\ref{eq:phidiscrevo}), and (\ref{eq:phidiscrtur}) into the form \citep{2010JOSAA..27.122M}
\begin{eqnarray}
\begin{array}{rcll}
\mathbf x_{n+1}&=&\mathsf A\mathbf x_n+\mathbf v_n&\qquad\,\,\,\  \qquad\qquad\textrm{[equation of state],}\end{array}\label{eq:statespace1}\\
\begin{array}{rcll}
y_n&=&\mathsf C\mathbf x_n-u_{n-2}+w_n&\qquad\textrm{[observation
equation].}
\end{array}\label{eq:statespace2}
\end{eqnarray}
where (for the sake of the example, we assume two vibration components, $\mathrm{vib}\,1$ and $\mathrm{vib}\,2$)
\begin{eqnarray}
\mathbf x_n^\top&=&\big(
\varphi^\mathrm{tur}_{n}\,\,\,\varphi^\mathrm{tur}_{n-1}\,\,\,
\varphi^\mathrm{vib\,1}_{n}\,\,\,\varphi^\mathrm{vib\,1}_{n-1}\,\,\,
\varphi^\mathrm{vib\,2}_{n}\,\,\, \varphi^\mathrm{vib\,2}_{n-1}\,\,\,
\big),\\\mathbf v^\top_n&=&\big(v^\mathrm{tur}_{n}\,\,\,0\,\,\,
v^\mathrm{vib\,1}_{n}\,\,\,0\,\,\,
v^\mathrm{vib\,2}_{n}\,\,\,0\big),\\
\mathsf C&=&\big(0\,\,\,1\,\,\,0\,\,\,1\,\,\,0\,\,\,1\big),
\end{eqnarray}
and
\begin{equation}
\mathsf A\,=\,
\begin{pmatrix}
a_1^\mathrm{tur}      &a_2^\mathrm{tur}&0&0&0&0\\    1&0&0&0&0&0\\
0&0& a_1^\mathrm{vib\,1}&a_2^\mathrm{vib\,1}&0&0\\0&0&1&0&0&0\\
0&0&0&0&a_1^\mathrm{vib\,2}&a_2^\mathrm{vib\,2}\\0&0&0&0&1&0\\
\end{pmatrix}.
\end{equation}
The vector $\mathbf x$ is called the state vector of the system. In addition, we denote by $\mathsf\Sigma_v$ and $\mathsf\Sigma_w=\sigma_w^2$ the covariance matrices of the system noise ($\mathbf v$) and measurement noise ($w$), respectively. We note that for the current example, $\mathsf\Sigma_v$ is of the form
\begin{equation}
\mathsf\Sigma_v\equiv\mathrm E\{\mathbf v\mathbf
v^\top\}=\mathrm{diag}\big(\sigma_v^{\mathrm{tur}},0,\sigma_v^{\mathrm{vib}\,1},0,\sigma_v^{\mathrm{vib}\,2},0\big)^2.
\end{equation}
As usual in state-space representation, the state variable $\mathbf x_n$ is a hidden quantity: only $y_n$ is observed. It is important to highlight the block-diagonal form of the matrix $\mathsf A$: every perturbation component (turbulence and vibrations) contributes a single AR(2) block.

\subsection{Control: the (asymptotic) Kalman filter}\label{sect:Kalman}
Kalman filtering \citep{1960TASME..82.35K} is a technique designed for general data processing. It is used to track and correct observed systems in a statistically optimal way, in that it minimizes the residual signal \citep{2009Kalman.book.....C}. Kalman filters start from a linear state-space description of the system and provide estimations of the system's (hidden) state vector derived from the observations.

We denote by $\hat{\mathbf x}_{n|n'}$ the estimation of the state vector at time step $n$, based on all observations up to time step $n'$ (i.e., $y_0,\ldots, y_{n'}$). Given the state-space model in Eqs.~(\ref{eq:statespace1}) and (\ref{eq:statespace2}), the Kalman filter equations\footnote{Note that Eqs.~(\ref{eq:kalmaneqs1}) and (\ref{eq:kalmaneqs2}) represent the asymptotic Kalman equations. The non-asymptotic equations can be found e.g.~in \citet{2009Kalman.book.....C}.} take the form
\begin{eqnarray}
\hat{\mathbf x}_{n|n}&=&\hat{\mathbf x}_{n|n-1}+\mathsf G_\infty(y_n-\mathsf C\hat{\mathbf x}_{n|n-1}+u_{n-2})\label{eq:kalmaneqs1},\\
\hat{\mathbf x}_{n+1|n}&=&\mathsf A\hat{\mathbf
x}_{n|n}.\label{eq:kalmaneqs2}
\end{eqnarray}
The former of these equations updates the information about the disturbance state $\mathbf x_n$, taking into account the measurement $y_n$ that has become available at time step $n$. The latter then predicts the state at the subsequent moment in time.

The matrix $\mathsf G_\infty$ in Eq.~(\ref{eq:kalmaneqs1}) is the (asymptotic) Kalman gain matrix, calculated as
\begin{equation}
\mathsf G_\infty=\mathsf\Sigma_\infty\mathsf C^\top(\mathsf
C\mathsf\Sigma_\infty\mathsf
C^\top+\mathsf\Sigma_w)^{-1},\label{eq:gain}
\end{equation}
where $\mathsf \Sigma_\infty$ is the solution of the matrix
(Riccati) equation
\begin{equation}
\mathsf\Sigma_\infty=\mathsf A\mathsf\Sigma_\infty\mathsf
A^\top-\mathsf A\mathsf\Sigma_\infty\mathsf C^\top(\mathsf
C\mathsf\Sigma_\infty\mathsf C^\top+\mathsf\Sigma_w)^{-1}\mathsf
C\mathsf\Sigma_\infty\mathsf
A^\top+\mathsf\Sigma_v.\label{eq:riccati}
\end{equation}
The gain matrix determines the weight by which the new measurement $y_n$ is taken into account for updating our estimate of the state vector $\mathbf x_n$. The mathematical form of $\mathsf G_\infty$ can be derived by requiring that the expectation value $\mathrm E\big\{(\mathbf x_n-\hat{\mathbf x}_{n|n})^2\big\}$ is minimal (e.g.~\citealt{2009Kalman.book.....C}). We note that the gain matrix contains the noise characteristics ($\mathsf\Sigma_v,\mathsf\Sigma_w$), making it optimally adapted to the considered system.

The final piece of information that is needed to design a two-telescope control scheme is an equation to determine commands. An optimal command will minimize the observed residual OPD $y_n$, which was given by Eq.~(\ref{eq:statespace2}). The command $u$ needs to compensate the $\mathbf x$-part ($w$ is the noisy part, which cannot be reduced by $u$). It is not difficult to see that the most appropriate choice of $u$ is
\begin{equation}
u_n=\mathsf{K}\,\hat{\mathbf x}_{n+1|n}, \label{eq:command}
\end{equation}
where (still for the two-vibration example)
\begin{equation}
\mathsf K=\big(1\,\,\,0\,\,\,1\,\,\,0\,\,\,1\,\,\,0\big).
\end{equation}
This result forms the final equation needed for control.

\section{Parameter identification}\label{sect:identification}
The state-space description that we need to model the fringe-tracking process contains a possibly large number of parameters. Reconsidering the state-space model in Eqs.~(\ref{eq:statespace1}) and (\ref{eq:statespace2}), we see that the parameters to identify are
\begin{equation}\big\{(a^i_1,a^i_2,\sigma^i_v)\,|\,\forall i\in\{\textrm{disturbances}\}\big\}\qquad\textrm{and}\qquad \sigma_w.\end{equation}
Once the parameters are obtained, we can ``fill'' the system matrices, and apply the Kalman filter.

We assume that OPD measurements from the phase sensor are the only available piece of information about the disturbances. In other words, it is assumed that no other \emph{a priori} information about the disturbance components is available (e.g.~from other observation techniques).

\begin{figure}[!t]
\centering
\includegraphics[width=0.49\textwidth,viewport=0 0 420 310,clip]{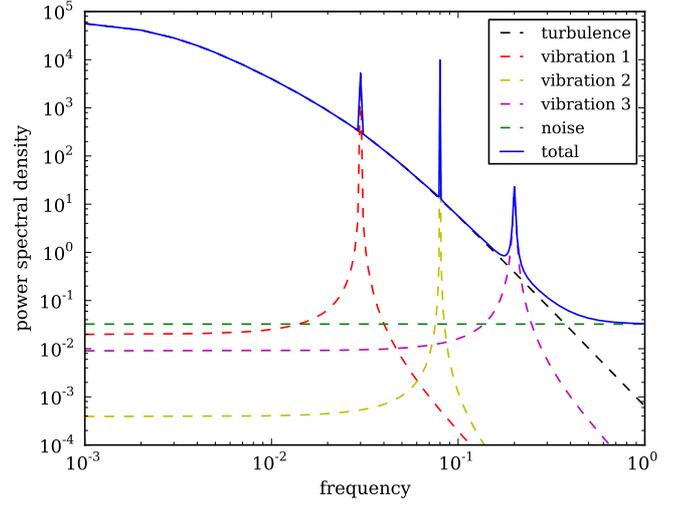}
\caption{A general perturbation power spectrum consists of a
turbulence spectrum, a flat (measurement) noise spectrum and vibration
peaks.}\label{fig:components}
\end{figure}

\subsection{The spectral method}
The method we use to calculate model parameters is based on the one proposed by \citeauthor{2010JOSAA..27.122M} (\citeyear{2010JOSAA..27.122M}, hereafter the MPFK method). The MPFK method is a power-spectrum-based routine designed for the parameter identification of the Kalman-filter AO control scheme of \citet{2004JOSAA..21.1261L} and \citet{2006PhDT.........2P}, the scheme that lies at the basis of Sect.~\ref{sect:Kalman2}.  We refer to the original MPFK paper for details of the method, and give here a brief summary and some information on our modifications.

The operation of the MPFK method is based on modeling the (estimated) power spectrum of an observed perturbation-phase sequence. In a time interval before identification, a sequence of closed-loop data is obtained and corrected for the applied commands (pseudo-open loop data, POL). In the notation of the state-space model, a sequence of data
\begin{equation}
\{y_n^\mathrm{POL}\equiv y_n+u_{n-2}\,\,|\,\,n=0,\ldots,N-1\}\label{eq:ypol}
\end{equation}
is collected. The periodogram of this sequence contains the imprint of the different perturbation components that contribute to the phase signal (Fig.~\ref{fig:components}). In the first fitting step, the measurement noise is estimated from the flat tail of the periodogram. By stepwise adding AR(2)-model spectra to the estimated noise spectrum, first for the turbulence and then for the vibration components, the global power spectrum is assembled until a stopping criterion is satisfied. The iterative fitting itself is based on the maximization of the likelihood function associated with each model periodogram. We note that the MPFK method is designed to operate in non-real time, i.e.~outside the real-time tracking loop.

In the standard MPFK method, a vibration-peak fitting step makes use of a large parameter grid for the three parameters associated with a single vibration component ($a^i_1,a^i_2,\sigma^i_v$). The method uses no \emph{a priori} information about the vibration components, but simply maximizes in each iteration the likelihood criterion corresponding to the given grid. In other words, the fitting of vibration peaks is done in a ``blind'' way on a large grid. To limit the computation time, we decided to add a simple subroutine that selects peaks from the power spectrum. In every iteration, we flag the points that differ significantly from the estimated power spectrum resulting from the previous iteration. Picking each time the highest peak allows us then to use a much smaller grid around a first guess for the peak parameters, which is made by estimating the maximum, width, and central frequency of the peak. In addition to the gain in computation time, peak picking limits the risk that multiple peaks are being fit with a single large peak, degrading the quality of the fit. We note that our peak-picking method selects peaks based on their height, rather than their energy.

\subsection{Practical application}
In the MPFK paper, it is shown that $N=2000$ is a proper choice for the length of the POL-data acquisition. Since POL data is constructed from closed-loop data, the acquisition of data sequences does not require the loop to be (re)opened. This property is advantageous for the operation of the fringe tracker and opens perspectives to perform updates of the model parameters in closed loop. There is however a fundamental difference between the case of AO, for which the MPFK method was designed originally, and interferometry. In the former case, the initialization of the Kalman filter can be based on a sequence of open-loop data (no POL data is available at initialization). In the case of fringe tracking, however, it is impossible to obtain a sequence of open-loop data without taking the risk that fringes are lost. Unlike the case of AO, where wavefront observables remain well-defined at the level of the wave-front sensor (e.g.~microlens images for a Shack-Hartmann wavefront sensor), uncorrected fringes have a typical root mean square (rms) motion that largely surpasses typical detection ranges.

The initialization of the Kalman-filter controller for fringe tracking thus inevitably requires a different strategy than open-loop measurements. A straightforward solution is to close the loop temporally using a classical control algorithm. POL data obtained in this way can then be fitted, and consequently we switch from the classical to the Kalman-filter controller. We note that this technique assumes that we take into account the difference in measurement error: the measurement noise $\sigma_w$ for a classical controller, extracted from the POL-data fit, will be higher than for the Kalman filter. The measurement error for the latter can be estimated based on Eq.~(\ref{eq:sigmawij}), which is introduced in Sect.~\ref{sect:simulations}.


\subsection{Other methods}
The spectral identification method is based on a maximum-likelihood method to model periodograms of measurement sequences. Alternative methods exist to identify parameters in state-space models (see also \citealt{2010JOSAA..27.122M}).

The advantage of the spectral method is that it is rather transparent, and gives satisfying results (see Sect.~\ref{sect:simulations}). The reason is that the imprint of the different perturbation components can be largely decoupled in frequency space, which makes it possible to separate the different disturbances. Future tests on real data will have to justify the use of the method for fringe-tracking purpose, and allow us to optimize the parameter-grid construction. A disadvantage of the method, however, is that it is difficult in general to make an iterative fitting routine robust against outliers and non-standard situations. A good alternative to the spectral-fit method could be a method of a purely algebraic nature, e.g.~similar to the one proposed by \citet{2005SPIE.5903...75H}. We are currently checking the possibility of using methods of this nature for parameter identification, as an alternative to the spectral method introduced here. Tests of identification methods based on the reconstruction of the measured time sequences \citep{1982JTSA..19.1519S,1993ITSP...41.3202Z} were not successful in the current case, basically owing to the large number of components that need to be characterized.

\section{Four-telescope fringe-tracking control}\label{sect:Kalman4}
In the two previous sections, we have presented a general and complete control scheme for two-telescope fringe tracking. It is generally known that two-telescope interferometers provide a rather limited amount of information per observation (one point in the $uv$-plane per wavelength bin, no direct phase information). Optical interferometry needs to be extended to include more apertures, especially when considering image reconstruction

The purpose of the current section is to find how two-telescope control can be extended to more telescopes. We chose to consider the case of four-telescope fringe tracking, in the framework of the second-generation near-infrared beam combiner GRAVITY for VLTI. GRAVITY will combine the beams of the four UTs (or four 1.8-m Auxiliary Telescopes, ATs) in a completely redundant way, i.e.~on six baselines. Per elementary fringe exposure, the instrument therefore measures six OPDs, rather than one OPD for a two-telescope beam combiner. The OPDs are corrected by a system of four actuators, i.e.~one per telescope. We now consider how the six observables can be transformed into proper commands for the actuator system. We note that this problem is absent in the case of a two-telescope beam combiner: one OPD measurement corresponds to one command. Our four-telescope approach can easily be generalized to $n$-telescope fringe tracking.

\subsection{General modal-based control}\label{sect:modalbasedcontrol}
Piston disturbances acting on a four-telescope interferometer can be considered as vectors in a four-dimensional space. In this space, an arbitrary disturbance $\mathbf P$ can be expressed as
\begin{equation}
\mathbf P=\big(P^0\,\,P^1\,\,P^2\,\,P^3\big)^\top.
\end{equation}
The actual piston disturbances $\mathbf P$ are not directly observable: only piston differences or optical path differences (OPDs) are measured. In a similar way to above, we can define OPD vectors as
\begin{equation}
\mathbf{OPD}=\big(\mathrm{OPD}^{01}\,\,\mathrm{OPD}^{02}\,\,\mathrm{OPD}^{03}\,\,\mathrm{OPD}^{12}\,\,\mathrm{OPD}^{13}\,\,\mathrm{OPD}^{23}\big)^\top,
\end{equation}
where $\mathrm{OPD}^{ij}\equiv P^j-P^i$. For simplicity, we have assumed that pistons and OPDs are expressed in the same units, say microns. The conversion $\mathbf P\rightarrow \mathbf{OPD}$ is then given by the matrix equation
\begin{equation}
\mathbf{OPD}=\mathsf M\,\mathbf P,\label{eq:OPDMP}\end{equation}
where
\begin{equation}\mathsf
M^\top=\begin{pmatrix}-1&-1&-1&0&0&0\\1&0&0&-1&-1&0\\0&1&0&1&0&-1\\0&0&1&0&1&1
\end{pmatrix}.\label{eq:Mmatrix}
\end{equation}

It is easy to show that the system in Eq.~(\ref{eq:OPDMP}) is not invertible. Mathematically, the matrix $\mathsf M$ is of rank three, rather than four. Physically, this corresponds to it being impossible to recover the global piston $P^\mathrm{tot}\equiv P^0+P^1+P^2+P^3$ from OPD observations (in a similar way to the piston not being observable with a single-dish telescope). A four-telescope fringe tracker therefore only needs to correct for three net observables.

\begin{figure}
\centering
\includegraphics[width=0.5\textwidth,viewport=120 300 470 530,clip]{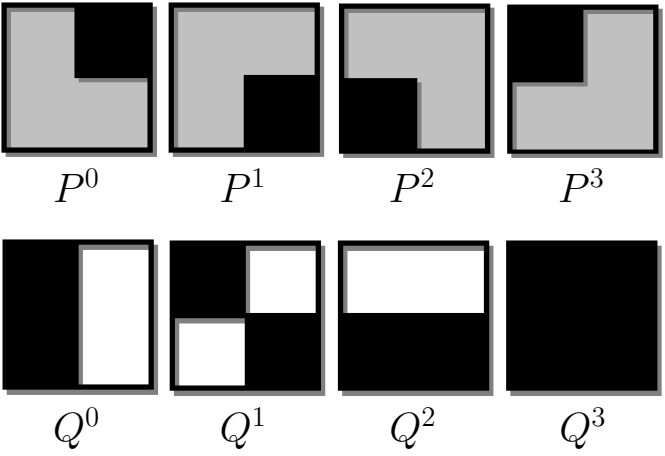}
\caption{Graphical representation of the four pistons (top) and the four modes (bottom). The four telescopes are represented as four pixels per square. Grey zones denote the zero-positions, black zones denote positive displacement, and white zones equal negative displacement. The modes could for example be labeled as tilt, twist, tip, and global piston. They form an equivalent to Zernike modes in AO, but for a four-telescope interferometer.}\label{fig:modes}
\end{figure}

Rather than using a description that is based on pistons, we perform a matrix transformation, based on a singular value decomposition of the matrix $\mathsf M$, to construct a set of variables that directly isolates the invisible global piston. We define the \emph{mode} four-vector $\mathbf Q$ ($Q^i$ for $i=0,1,2,3$) as
\begin{equation}
\mathbf Q\equiv \mathsf V^\top\,\mathbf P,\label{eq:mode}
\end{equation}
where
\begin{equation}
\mathsf V=\frac12\begin{pmatrix} -1&-1&-1&1\\
-1&1&1&1\\
1&-1&1&1\\
1&1&-1&1
\end{pmatrix}.
\end{equation}
This orthogonal transformation isolates (a multiple of) $P^\mathrm{tot}$ into $Q^3$. The Q-modes form an equivalent to the Zernike modes in AO, and their relation to the pistons is graphically represented in Fig.~\ref{fig:modes}. The conversion between vectors $\mathbf{OPD}$ and $\mathbf{Q}$ is then given by
\begin{equation}
\mathbf Q=\mathsf H\,\,\mathbf{OPD},\label{eq:SdagUTi}\end{equation}
where
\begin{equation}\mathsf H= \frac14\begin{pmatrix}
0&1&1&1&1&0\\1&0&1&-1&0&1\\1&1&0&0&-1&-1\\0&0&0&0&0&0
\end{pmatrix}.\label{eq:Hmatrix}\end{equation}
We note that the matrix $\mathsf H$ automatically puts $Q^3$ to 0, indicating again that we have no information about $P^\mathrm{tot}$ from vectors $\mathbf{OPD}$. The three components $Q^0$, $Q^1$, and $Q^2$ now form a proper description of the observable system with three degrees of freedom. A general control scheme based on modes consists then of three steps:
\begin{enumerate}
\item We calculate the residual mode vector $\mathbf y_Q$ from the measured residual OPD 6-vector $\mathbf y$ using the transformation in Eq.~(\ref{eq:SdagUTi})
\begin{equation}
\mathbf y_Q=\mathsf H\,\mathbf y.\label{eq:SdagUT}
\end{equation}
\item Using three controllers (e.g.~integrator controllers), we calculate a proper Q-command $\mathbf u_Q$ to correct the Q-modes. We fix $u_{Q}^3\equiv0$, that is, the global piston is fixed (by convention).
\item Transform $\mathbf u_Q$ into the four-component piston command $\mathbf u$ using the inverse transformation of Eq.~(\ref{eq:mode}): $\mathbf u=\mathsf V\,\mathbf u_Q$. These four commands are sent to the actuators for piston correction.
\end{enumerate}

\subsection{Modal Kalman-filter control}\label{sect:modalKalmanfilter}
Above, we introduced a properly defined control scheme based on a modal approach. We now apply a similar philosophy to a Kalman-filter based control model, and find the first appropriate control law for four-telescope fringe tracking.

\subsubsection{Modal state-space
model}\label{sect:modalstatespacemodelmain} Starting from the state-space description in Sect.~\ref{sect:Kalman}, we derived (see Appendix \ref{appendix:modalKalman}) a full modal state-space model for a four-telescope fringe-tracking system, which can be easily generalized to $n$ telescopes. Individual OPD measurements are assumed to be uncoupled in this model. Additionally, all piston perturbations are considered as being independent. We note that this condition is imposed on two scales. First, on the scale of one telescope, we ignore the coupling between different vibration components. Secondly, on the scale of all telescopes, we assume neither that coupling exists between different telescope vibrations, nor that the turbulence perturbation on the pistons is coupled\footnote{On scales comparable to or larger than the turbulence outer scale, however, it is true that this assumption will break down.}.

The result of this analysis is the modal state-space model (compare with Eqs.~(\ref{eq:statespace1}) and (\ref{eq:statespace2})) given by
\begin{eqnarray}
\mathbf x_{Q,n+1}&=&\widetilde{\mathsf A}\,\mathbf x_{Q,n}+ \mathbf v_{Q,n},\label{eq:eosmodemain}\\
\mathbf y_{Q,n}&=&\widetilde{\mathsf C}\,\mathbf x_{Q,n}-\mathbf
u_{Q,n-2}+\mathbf w_{Q,n}\label{eq:oemodemain}.
\end{eqnarray}
The vector quantities have the following definitions:\newline\newline
\begin{tabular}{rl}
$\mathbf x_Q$:&4-block modal state vector\\
$\mathbf y_Q$:&modal measurement 4-vector ($y_Q^3\equiv0$)\\
$\mathbf u_Q$:&modal command 4-vector ($u_Q^3\equiv0$)\\
$\mathbf v_Q$:&4-block modal system noise vector, of covariance\\&
matrix
$\mathsf\Sigma_{Q,v}$\\
$\mathbf w_Q$:&measurement noise 4-vector, of covariance
$\mathsf\Sigma_{Q,w}$; \\&note that $w_Q^3\equiv0$\\
\end{tabular}\newline\newline
On the other hand, the matrices $\widetilde{\mathsf A}$ and $\widetilde{\mathsf C}$ are
\begin{equation}
\widetilde{\mathsf A}=\begin{pmatrix} \mathsf A&\mathsf 0&\mathsf
0&\mathsf 0\\\mathsf 0&\mathsf A&\mathsf 0&\mathsf 0\\\mathsf
0&\mathsf 0&\mathsf A&\mathsf 0\\\mathsf 0&\mathsf 0&\mathsf
0&\mathsf A
\end{pmatrix}\qquad\textrm{and}\qquad\widetilde{\mathsf C}=\begin{pmatrix} \mathsf C&\mathsf 0&\mathsf
0&\mathsf 0\\\mathsf 0&\mathsf C&\mathsf 0&\mathsf 0\\\mathsf
0&\mathsf 0&\mathsf C&\mathsf 0\\\mathsf 0&\mathsf 0&\mathsf
0&\mathsf 0
\end{pmatrix},\label{eq:wideAwideC}
\end{equation}
where the blocks $\mathsf A$ and $\mathsf C$ have the same structure as the corresponding system matrices in the two-telescope state-space model. All scalar quantities ($y_n$, $u_n$, $w_n$) have now become four-vectors, and the vector/matrix quantities ($\mathbf x_n$, $\mathsf A$, etc.) have obtained a four-block structure ($\mathbf x_{Q,n}$, $\widetilde{\mathsf A}$, etc.). This is all because there are four modes. There are several interesting properties about the above result, which we discuss now. We note that we first neglect the noise terms.

First, all four terms in Eq.~(\ref{eq:oemodemain}) have a last component that is immediately defined to be 0 (see also matrix $\widetilde{\mathsf C}$). In other words, the global piston is automatically defined as not being observable. Thus, although it is still a real evolving quantity described by the last block component of the vectors in Eq.~(\ref{eq:eosmodemain}), we can neglect it for the analysis.

Secondly, when looking at the structure of the state-space model and of Eq.~(\ref{eq:wideAwideC}), we see that all modes have a decoupled evolution (still neglecting the noise). What is more is that the sub-matrices of $\widetilde{\mathsf A}$ are the \emph{same}, and the same observation holds for $\widetilde{\mathsf C}$. Neglecting thus the noise, we see that the evolution is described by four identical copies of a two-telescope state-space model (with the modification that for the last copy the observation equation is 0). Modes therefore essentially behave in a similar way to the quantity $\varphi$ of the previous sections.\footnote{Applying the strategy described in Appendix \ref{appendix:modalKalman} to the two-telescope case shows that there are two modes: $P^1-P^0$ and $P^0+P^1$. The former of these \emph{is} exactly (a multiple of) $\varphi$; the state-space model in Sect.~\ref{sect:Kalman2} is already in a modal form.}

\paragraph{Considering the noise terms.} A few things change when  we also take into account the noise terms. Since modes are calculated  as combinations of pistons, Eq.~(\ref{eq:mode}), or combinations of  OPDs, Eq.~(\ref{eq:SdagUT}), the covariance matrices are in general no longer diagonal (given our assumption that piston perturbations as  well as OPD measurements are independent).

Since we have individual OPD measurements at our disposal, it is possible to estimate the measurement errors $\sigma^{ij}_w$ associated with the measurements of $\mathrm{OPD}^{ij}$. This allows us to calculate the mode measurement error covariance matrix $\mathsf\Sigma_{Q,w}$ as
\begin{equation}\mathsf\Sigma_{Q,w}=\mathsf
H\,\mathsf\Sigma_w\,\mathsf H^\top,\label{eq:SigmaQw}
\end{equation}
where
\begin{equation}
\mathsf\Sigma_w=\mathrm{diag}\big(\sigma^{01}_w,\sigma^{02}_w,\sigma^{03}_w,\sigma^{12}_w,\sigma^{13}_w,\sigma^{23}_w\big)^2
\end{equation}
(we refer to Appendix \ref{appendix:modalKalman}). The evaluation of the covariance matrix $\mathsf\Sigma_{Q,v}$, however, is a more fundamental problem. The matrix $\mathsf\Sigma_{Q,v}$ depends on quantities that require direct piston information, which is inaccessible (owing to the unobservability of the absolute pistons). We neglect the off-diagonal entries; under this approximation, it can be shown that  $\mathsf\Sigma_{Q,v}$ has a diagonal built of four equal diagonal parts (e.g.~like the structure of $\widetilde{\mathsf A}$). In this model, the evolution of the four modes by Eq.~(\ref{eq:eosmodemain}) is thus genuinely independent and equal.

\paragraph{Convention.} Up to now, we have carried the global piston  mode as a superfluous burden in our matrix equations. Since we assume that the modes are fully decoupled, we can drop all  matrix and vector blocks corresponding to $Q^3$. Under this convention, the fundamental matrices $\mathsf V$ and $\mathsf H$ become
\begin{equation}
\mathsf V=\frac12\begin{pmatrix} -1&-1&-1\\
-1&1&1\\
1&-1&1\\
1&1&-1
\end{pmatrix}\quad\textrm{and}\quad\mathsf
H= \frac14\begin{pmatrix} 0&1&1&1&1&0\\1&0&1&-1&0&1\\1&1&0&0&-1&-1
\end{pmatrix}.\end{equation} 
It is not difficult to generalize this operation to the other quantities. For example, the measurement-error covariance matrix $\mathsf\Sigma_{Q,w}$ loses its last column and line and becomes a $3\times3$ matrix.\footnote{Using Eq.~(\ref{eq:SigmaQw}), it can be shown that the last column and row were zeros anyway, exactly owing to their unobservability.}

\subsubsection{Modal Kalman filter}
The non-diagonal nature of the covariance matrix $\mathsf\Sigma_{Q,w}$ shows that the modes are not independently measured. It follows that the control scheme has to involve one global Kalman filter for the three modes, rather than one per mode. The full Kalman-filter equations then take the form
\begin{eqnarray}
\hat{\mathbf x}_{Q,n|n}&=&\hat{\mathbf x}_{Q, n|n-1}+ \mathsf
G_\infty\,(\mathbf y_{Q,n}-\widetilde{\mathsf C}\,
\hat{\mathbf x}_{Q,n|n-1}+\mathbf u_{Q,n-2}),\\
\hat{\mathbf x}_{Q,n+1|n}&=&\widetilde{\mathsf A}\,\hat{\mathbf
x}_{Q,n|n}.
\end{eqnarray}
In this equation, the Kalman gain $\mathsf G_\infty$ is calculated as in Eqs.~(\ref{eq:gain}) and (\ref{eq:riccati}), with the changes $\mathsf A, \mathsf C, \mathsf\Sigma_v, \mathsf\Sigma_w\rightarrow\widetilde{\mathsf A}, \widetilde{\mathsf C}, \mathsf\Sigma_{Q,v}, \mathsf\Sigma_{Q,w}$. Finally, the optimal mode command is calculated as
\begin{equation}
\mathbf u_{Q,n}=\widetilde{\mathsf K}\,\hat{\mathbf x}_{Q,n+1|n},
\end{equation}
where the matrix $\widetilde{\mathsf K}$ is defined in the same way as $\widetilde{\mathsf C}$ in Eq.~(\ref{eq:wideAwideC}), with blocks $\mathsf K$ defined as in Sect.~\ref{sect:Kalman}.

\subsubsection{Practical considerations}
The modal Kalman-filter control is an interesting control scheme: it is completely symmetric, its conversion matrices are simple, its input and output have the same dimensions (three), and (under the above approximations) each mode has exactly the same state-space model. An additional advantage of the modal scheme is presented in Sect.~\ref{sect:4telid}, where we consider the identification of the parameters in the state-space model described in Eqs.~(\ref{eq:eosmodemain}) and (\ref{eq:oemodemain}).

The symmetry involved in the definition of the modes implies that the control will work most effectively when all other involved quantities are symmetric. It is easy to see, however, that this is not the case. The design and implementation of the GRAVITY system are largely symmetric, whereas actual observation conditions (e.g.~wind shake, vibrations) and baseline configurations are not. Whenever a single component in the system fails, for example one telescope or one baseline, at least two of the three observables are spoiled in the current control scheme (each OPD contributes to two modes). \emph{A priori}, the loss of observables might be expected to have a negative impact on the control. We therefore propose a second four-telescope control scheme, which is based on the tracking of the individual OPDs.

\subsection{OPD Kalman-filter control}\label{sect:OPDKalman}
An alternative to a global Kalman filter that tracks all modes is to use six Kalman filters to control each individual OPD. In this (redundant) scheme, each telescope pair is considered as a single two-telescope interferometer.

One might wonder why it would be more interesting to track twice the number of variables (three modes versus six OPDs). The main interest of the \emph{modal control} is that it was defined in a perfectly symmetric way to transform the six observables and four pistons into three quantities that are tracked. Here, the conversion to and from modes is done using fixed matrices. In the \emph{OPD control}, however, the redundancy of the conversion to and from OPDs allows us to choose different ways of defining commands. In particular, one can consider non-symmetric combinations of observables or---even better---adaptively weight observables with respect to the physical conditions of observations. The latter possibility is our main point of interest: whenever observing conditions (low flux, low-signal baselines, etc.) make certain OPDs unfavorable, we can lower their weight in the command calculation. Additionally, it even allows us to let the interferometer work with only three or two telescopes.

We postpone a deeper discussion of the above problem to Sect.~\ref{sect:viszero}, where we also modify the modal scheme to a more robust version. For completeness, we already give an appropriate method for command-vector calculation. By $\mathbf x_\textrm{OPD}$, we denote the six-block OPD state vector, i.e.~the state vector corresponding to each OPD organized into a column vector. The four-component piston command $\mathbf u_n$ is then calculated as
\begin{equation}
\mathbf u_n=\mathsf M^\dag_W\,\widetilde{\mathsf K}\,\hat{\mathbf
x}_{\mathrm{OPD},n+1|n},
\end{equation}
where $\widetilde{\mathsf K}$ is the six-block version of the previously introduced matrix $\mathsf K$, in order to operate on the six substate vectors in $\hat{\mathbf x}_\textrm{OPD}$. The matrix $\mathsf M^\dag_W$, on the other hand, is a weighted generalized inverse to the matrix $\mathsf M$ (Eq.~(\ref{eq:Mmatrix})), calculated as
\begin{equation}
\mathsf M^\dag_{W}=(\mathsf M^\top\mathsf W\,\mathsf M)^\dag\,\mathsf M^\top\mathsf W.\label{eq:MWmatrix}
\end{equation}
In this equation, the $6\times6$ weighting matrix $\mathsf W$, defined further on in Eq.~(\ref{eq:Wmatrix}), distributes the weight among the different OPD substates for the command calculation.

\subsection{Four-telescope parameter identification}\label{sect:4telid}
The MPFK method (including the peak-pick extension) can be directly applied to two-telescope fringe tracking, for the simple reason that it is exactly designed for the corresponding state-space model. In the four-telescope case, the sequences of $y_n^\mathrm{POL}$ (Eq.~(\ref{eq:ypol})) are replaced by equivalent sequences of six-dimensional vectors $\mathbf y_n^\mathrm{POL}$, i.e.~the POL OPD measurement six-vectors.

In the modal state-space model, the three observed modes are decomposed into the elementary disturbances. Therefore, the most logical option is to apply the spectral fit to the time sequences
\begin{equation}
\big\{\mathbf y^\mathrm{POL}_{Q,n}\equiv \mathsf H\mathbf
y^\mathrm{POL}_n\,|\, n=0,\ldots,N-1\big\},
\end{equation}
where $\mathbf y^\mathrm{POL}_{Q,n}$ are the three-component POL mode measurements (we use Eq.~(\ref{eq:SdagUT})). Under the convention of neglecting the off-diagonal entries of the covariance matrix $\mathsf\Sigma_{Q,v}$, we have shown in Sect.~\ref{sect:modalstatespacemodelmain} that the state-space model is the same for the three modes. The result of this approximation is not only that we can apply the spectral fit to each of the three non-trivial modes, but also that we can average the spectra of the three modes and perform one global fit. Indeed, assuming three independent modes, the three periodograms are independent realizations of the same state-space model. Averaging periodograms for identification translates into a modification of the periodogram statistics, and thus a new likelihood function for the identification method (see Appendix \ref{appendix:details} for some details). The important advantage is that the periodogram is a more accurate estimation of the power spectrum, and that even small perturbation components become significantly discernible from statistical noise.

The OPD Kalman-filter control scheme, on the other hand (Sect.~\ref{sect:OPDKalman}), is based on the redundant tracking of all OPDs by six two-telescope Kalman filters. In this case, it thus suffices to apply the spectral fit to each of the six OPD measurement sequences. The disadvantage is, of course, that we have to apply the identification method six times, instead of once in the modal scheme. Additionally, since we have only one sequence per OPD, we cannot apply the advantageous averaging of periodograms.

\section{Fine-tuning the fringe-tracking control}\label{sect:viszero}
Our aim to find appropriate Kalman-filter control schemes led us to consider two options: tracking three observable modes in a three-component Kalman filter and tracking six OPDs in six individual Kalman filters. We now consider some specific cases encountered in real observations, and present a way of dealing with them in our control schemes.

\subsection{Low S/N baselines}
Different processes can lead to lower-quality OPD measurements on a given baseline. Examples are non-central fringe measurements in the fringe envelope and variable splitting ratios of the beams. Whenever the fringe amplitude on a certain baseline is low with respect to the noise level, the control strategy should be designed to maintain the fringe tracking at as optimal a level as possible.

An advantage of the interferometer concept considered above is that its redundant architecture enables it to take into account low fringe-amplitude baselines, in which different OPD combinations can compensate for other OPD measurements (e.g.~$\mathrm{OPD}^{01}=\mathrm{OPD}^{02}-\mathrm{OPD}^{12}$). The key step will be to take a weighted recombination of the six baselines, in order to mutually improve the raw measurements.

In essence, we can write the process of weighting the residual OPD values as
\begin{equation}
\mathbf{y}_W=\mathsf I_W\,\mathbf{y},\qquad\textrm{where}\qquad\mathsf I_W=\mathsf
M\,\mathsf M^\dag_W,\label{eq:weightedOPD}
\end{equation}
with an associated measurement-error covariance matrix
\begin{equation}
\mathsf\Sigma_{w,W}=\mathsf I_W
\,\mathsf\Sigma_w\,\mathsf I_W^\top.\label{eq:weightedOPDsig}
\end{equation}
The $6\times6$ matrix $\mathsf I_W$ combines in a weighted way the OPD measurements, which allows us to mutually improve each OPD value. The weights $W^{ij}$ attributed to each OPD$^{ij}$ are organized in a diagonal weighting matrix $\mathsf W$, which is used to calculate the previously introduced matrix $\mathsf M_W^\dag$ (Eq.~(\ref{eq:MWmatrix})). In accordance with the purpose of weighting, a proper definition of $\mathsf W$ is 
\begin{equation}
\mathsf W =\mathsf\Sigma_w^{-1},\label{eq:Wmatrix}
\end{equation}
which is the inverse of the (diagonal) measurement-error covariance matrix. In other words, we take
$W^{ij}=(\sigma^{ij}_w)^{-2}$. We note that either global weights $\mathsf W$, i.e.~fixed weights during operation, or instantaneous weights $\mathsf W_n$, i.e.~defined for each OPD measurement, can be considered. The latter is possible owing to an instantaneous measurement error that is provided by the phase sensor (we refer to Eq.~(\ref{eq:sigmawij}) in Sect.~\ref{sect:simulations}).

In the modal Kalman-filter control scheme, we now simply apply the standard OPD-to-mode transformation in Eq.~(\ref{eq:SdagUT}) to the weighted modes $\mathbf y_W$, and accordingly use a weighted measurement covariance matrix
\begin{equation}
\mathsf\Sigma_{Q,w,W}=\mathsf H\,\mathsf\Sigma_{w,W}\,\mathsf
H^\top.\label{eq:SigQwW}
\end{equation}
For the OPD scheme, the six Kalman filters are applied to the six weighted OPDs. We note that in this scheme, the weighting is assumed not to introduce coupling into the OPDs. In other words, we ignore off-diagonal terms in the covariance matrix $\mathsf\Sigma_{w,W}$ in the OPD scheme. This allows us to keep using the 6 independent Kalman filters, and will prove useful for the adaptive-gain definition below.

A final note on this OPD weighting is that it is only useful for non-resolved fringe-tracking targets. When the target is (significantly) resolved, the intrinsic visibility phases may be non-zero and hence phase closures cannot be assumed to be 0 (e.g.~$\mathrm{OPD}^{01}+\mathrm{OPD}^{12}- \mathrm{OPD}^{02}\ne0$, in general). In any case, fringe tracking of significantly resolved objects is disadvantageous owing to the lower fringe contrasts.

\subsection{Flux dropouts}\label{sect:fluxdropouts}
OPD weighting resolves the issue of low S/N's on individual baselines.  A second problem occurs when the flux acquired at a telescope is low, i.e.~the so-called flux dropouts. The notion of flux dropouts contains the different processes by which the number of photons coming from a certain telescope drops. These include:
\begin{itemize}
\item Atmosphere-related effects, where scintillation and other turbulence effects can lead to a failure of the AO system. The resulting Strehl ratio can be insufficient for proper fringe formation and detection. In addition, cirrus may lower the flux arriving at the focal plane of a telescope.
\item Errors in the fiber injection, where mechanical problems (e.g.~vibrations) and other propagating tip-tilt errors in the optical train can lead to problems in the injection of the beams into the fibers, coupled to the beam combiner.
\end{itemize}
As far as the second point is concerned, GRAVITY will have a fiber coupler subsystem to perform an internal tip-tilt correction \citep{2010SPIE.7734E..70P}. Owing to the residual errors, however, a specially developed fringe-tracking strategy is still required.

It should be clear that the problem of flux dropouts is very different from the problem introduced in the previous section. When the flux drops, it is impossible to compensate for the low signals on the baselines associated with that telescope: all these baselines are affected in a similar way. For non-predictive control schemes, no other solution exists apart from waiting until the lost signal re-emerges, while keeping the command for that telescope fixed. The advantage of using a \emph{predictive} method, such as a Kalman filter, is that it allows us to continue operating based on the state-space model.

In Appendix \ref{appendix:instgain}, we argue for the necessity to modify the gain to take into account flux problems at the level of a telescope. In essence, we modify the constant Kalman gain and allow it to decrease when observing conditions on associated telescopes degrade. The result is that the estimation of the next state is based more on the model, i.e.~the observed (noisy) residual OPD is not taken into account as effectively.

To introduce the instantaneous-gain definitions, we first define the instantaneous versions $\mathsf\Sigma_{Q,w,W,n}$ and $\mathsf\Sigma_{w,W,n}$ of the previously introduced noise-covariance matrices (Eqs.~(\ref{eq:weightedOPDsig}) and (\ref{eq:SigQwW})). The instantaneous covariances are recalculated at each time step using instantaneous weighting matrices $\mathsf W_n$ (rather than a constant $\mathsf W$). The discussion in Appendix \ref{appendix:instgain} then leads us to the instantaneous-gain definitions
\begin{eqnarray}
\mathsf
G_{\infty,n}=\frac{\mathrm{tr}\big\{\mathsf\Sigma_{Q,w,W}\big\}}
{\mathrm{tr}\big\{\mathsf\Sigma_{Q,w,W,n}\big\}}\,\mathsf
G_\infty\qquad[\textrm{modal scheme}]\label{eq:Ginftynmode}
\end{eqnarray}
(``tr'' denotes taking the trace) and
\begin{equation}
\mathsf
G_{\infty,n}^{ij}=\frac{\big(\mathsf\Sigma_{w,W}\big)^{ij}}{\big(\mathsf\Sigma_{w,W,n}\big)^{ij}}\mathsf
G_{\infty}^{ij}\qquad[\textrm{OPD scheme}]\label{eq:GinftyijnOPD}
\end{equation}
(the index $ij$ to the parentheses indicates that we take the matrix
diagonal element corresponding to the considered OPD$^{ij}$; for instance, for $ij=01$ and $ij=12$, we take the zeroth and third diagonal
element, respectively). 

The advantage of the OPD scheme is now clear. Here we consider the extreme case when we lose all information provided by one telescope. In the modal scheme, the loss of information immediately means that we lose all mode observables, which implies that we cannot take into account the observational information at that step (the instantaneous gain matrix becomes zero). On the other hand, losing a telescope in the OPD scheme means we still have three OPDs that are properly measured, and the gain for those Kalman filters can stay fixed. This implies that the command is still based on half the number of usual observables (versus none for the modal scheme). The cost we have to pay is that we need to track six variables instead of three.

\subsection{Piston isolation}
For some specific reasons, it might be interesting to completely decouple one or more telescopes from the command estimate. This can occur when a telescope (or its AO system) temporally becomes very unreliable during an observing run. The flexibility of the OPD scheme allows us to track blindly on one/two telescope(s) without affecting the tracking of the others, after which we can switch back to the usual scheme.

To temporally decouple telescope $i$, we attribute a weight zero to all associated baselines to telescope $i$. Doing so, however, results in putting the command on telescope $i$ to 0. This can be disadvantageous to the stability of the system, as it generally involves a large jump in all actuator positions (the average actuator position is always 0, hence all actuator positions change when one command is put to 0). The basic idea is then to add an extra component to the command vector,
\begin{equation}
\mathbf u_n=\mathsf M^\dag_W\,\widetilde{\mathsf K}\,\hat{\mathbf
x}_{\mathrm{OPD},n+1|n}+\mathsf L^i\,\widetilde{\mathsf
K}\,\hat{\mathbf x}_{\mathrm{OPD},n+1|n}.
\end{equation}
where the components of $\hat{\mathbf x}_{\mathrm{OPD},n+1|n}$ that correspond to $i$ are blindly estimated (gain 0). The newly-introduced matrix $\mathsf L^i$, on the other hand, has a specific structure that adds a blindly estimated command to $i$, and subtracts an equal value from all other commands such that the sum of the commands is zero. For instance, for $i=0$ (i.e.~to isolate $P^0$) and $i=\{1,2\}$ (i.e.~to isolate both $P^1$ and $P^2$), we have
\begin{equation}\mathsf L^0=\frac13\begin{pmatrix}
3&0&0&0\\-1&0&0&0\\-1&0&0&0\\-1&0&0&0
\end{pmatrix}\qquad\textrm{and}\qquad \mathsf L^{1,2}=\frac12\begin{pmatrix}
0&-1&-1&0\\0&2&0&0\\0&0&2&0\\0&-1&-1&0
\end{pmatrix},\end{equation}
respectively. As we subtract an equal command from each non-decoupled actuator, these commands on the decoupled telescope(s) do not affect the non-decoupled telescopes.

\section{Full Kalman-filter operation}\label{sect:full}
We have introduced the basic theory, developed several control schemes, and considered strategies that allow us to minimize the problems associated with imperfections of the system/control schemes. It is a good point now to define the full Kalman-filter operation by comparing the two schemes. The operation is split into an identification phase and a (real-time) tracking phase, the former being needed to find the parameters.

\paragraph{Identification phase.}\mbox{}\\
\begin{enumerate}
\item Pseudo-open loop (POL) data acquisition: a set of data points and estimated measurement errors is obtained in closed loop and corrected for the applied commands to get a POL sequence.
\item Weighting: we estimate a global measurement noise for each POL OPD sequence by taking the median of the measurement errors. We apply the weighting in Eq.~(\ref{eq:weightedOPD}) to get the weighted OPD sequence. In addition, for the modes, apply the OPD-to-mode transformation in Eq.~(\ref{eq:SdagUT}) to calculate the weighted mode sequence. We also calculate the weighted covariance matrices using Eq.~(\ref{eq:weightedOPDsig}) or (\ref{eq:SigQwW}).
\item Spectral fit: the spectral-fit method is applied to either the averaged mode periodograms (modal scheme) or the individual weighted OPD periodograms (OPD scheme) to get the fit parameters.
\item Filling of the Kalman matrices: the fitted parameters are used to define the system matrices, and to calculate the gain $\mathsf G_\infty$.
\end{enumerate}

\paragraph{Tracking phase.}\mbox{}\\
For each elementary fringe exposure:
\begin{enumerate}
\item Weighting: the residual OPD estimates for the six baselines, obtained from the phase sensor, are combined in a weighted way to give optimal residual mode estimates (modal scheme) or OPD estimates (OPD scheme). The weighting factors are calculated from instantaneous estimates of the measurement errors. We note that this step requires a singular-value decomposition to calculate the weighted generalized inverse $\mathsf M^\dag_W$ of $\mathsf M$. We calculate the associated instantaneous measurement covariances $\mathsf\Sigma_{Q,w,W,n}$ or $\mathsf\Sigma_{w,W,n}$ (Eq.~(\ref{eq:weightedOPDsig}) or (\ref{eq:SigQwW}) with instantaneous weights).
\item Kalman step: we apply the Kalman filter, using the instantaneous gain definitions (Eq.~(\ref{eq:Ginftynmode}) or (\ref{eq:GinftyijnOPD})).
\item Command calculation: we calculate the actuator commands. 
\end{enumerate}

\section{Simulations for VLTI/GRAVITY}\label{sect:simulations}
To test the performance of our Kalman-filter control, we construct a simple simulator of the control process in a fringe tracker, which also allows us to distinguish between the different control schemes. The simulations are performed in function of the key science case of the four-telescope beam combiner GRAVITY: the astrometric observation of the Galactic-center neighborhood. As described in \citet{2010SPIE.7734E..28G}, stable fringe tracking of a magnitude $\mathrm K=10$ star is required, with a specification of $300\,$nm for the residual OPD. 

More complete simulations of the fringe tracking process for GRAVITY, including frame-by-frame acquisition, are postponed to a future paper (\citealt{2012...C}, in preparation). In the same paper, a comparative study with classical control algorithms will be presented and loop parameters will be optimized. 

\begin{figure}[!t]
\centering
\includegraphics[width=0.5\textwidth, viewport=120 290 470 530, clip]{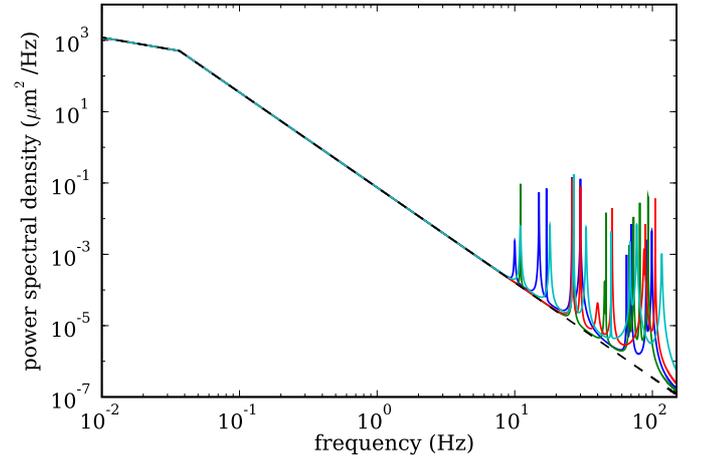}
\caption{Power spectra used to simulate the perturbations on the four
pistons (four colors). The turbulence-only spectrum is indicated by the
dashed line.}\label{fig:profile}
\end{figure}

\subsection{Disturbance simulation}\label{sect:disturbancesimulation}
Data is simulated based on model power spectra for the associated phenomena. Since turbulence and vibration components are assumed to be mutually independent, we can sum individual perturbation sequences to assemble complete data sequences. 

For each data sequence, we take a number of vibration components under the form of randomly excited damped oscillators. The frequencies chosen are inspired by OPD power spectra obtained with the VLTI/PRIMA fringe tracker, shown in \citet{2009A&A...507.1739S}. The total rms deviation per piston is chosen between $200$ and $240\,$nm, which is a realistic estimate of the current vibration level at VLTI \citep{2010SPIE.7734E.101P}. The vibration profiles used for each telescope are shown in Fig.~\ref{fig:profile} (superimposed on the turbulence profile), with some specifications in Table \ref{table:simulpar}.

Considering the turbulence contribution, \citet{1995JOSAA..12.1559C} derive the asymptotic laws for single turbulence-layer OPD spectra
\begin{equation}
S(f)\propto\left\{
\begin{array}{ll}
f^{-2/3}&\textrm{for}\quad0.2\,v/B>f,\\
f^{-8/3}&\textrm{for}\quad0.2\,v/B< f<0.3\,v/D,\\
f^{-17/3}&\textrm{for}\quad 0.3\,v/D<f
\end{array}\right.\label{eq:theoturbspec}
\end{equation}
(pure Kolmogorov turbulence, i.e.~with infinite outer scale), where $v$ is the wind speed parallel to the baseline, $B$ is the baseline, and $D$ is the diameter of the telescopes. The first two power laws in Eq.~(\ref{eq:theoturbspec}) are clearly observable in OPD spectra obtained with the Mark~III interferometer \citep{1987ApOpt..26.4106C}. In this case, the steep $f^{-17/3}$ cut-off is not observed because of to the small telescope diameters (high cut-off frequency). Nevertheless, the $f^{-17/3}$ law, which is also expected for mono-layer tip-tilt spectra, is also in practice not observed on large telescopes. The tip-tilt power spectra of the AO-system VLT/NAOS, used by \citet{2010JOSAA..27.122M} to verify the spectral identification method (MPFK method, Sect.~\ref{sect:OPDKalman}), are well-modeled by an AR(2)-model spectrum (lacking a very steep cut-off). In addition, OPD spectra obtained with the VLTI/PRIMA fringe tracker using the ATs ($D=1.8\,$m) are compatible with an $f^{-8/3}$ power law up to 200\,Hz, where a flat noise tail takes over \citep{2009A&A...507.1739S}. We thus conclude that, in real-life situations (multi-layer turbulence, integrated $C^2_N$-profiles, multi-directional wind), the $f^{-17/3}$ law is not observable/observed.

We therefore decide to use the OPD-spectrum model in Eq.~(\ref{eq:theoturbspec}), but discard the $f^{-17/3}$ cut-off. A very similar model is used by \citet{2010VLT..G} in the context of VLTI instrumentation. In Table \ref{table:simulpar}, we present the values assigned to the parameters, which are representative of the observing conditions at the VLTI. Each piston sequence is scaled to an rms deviation of 10\,$\upmu$m, such that the resulting OPD sequence has an rms value of $\sim14\,\upmu$m ($10\sqrt2\,\upmu$m). The theoretical turbulence profile is shown in Fig.~\ref{fig:profile}.

\subsection{Flux dropouts and measurement noise}\label{sect:fluxdropmeasnoise}
In addition, we need to introduce the phenomenon of varying flux in the simulations, which manifests itself as variable measurement noise. This means that, in parallel to the simulated OPD sequences, we need to simulate a time sequence of instantaneous noise characteristics, and add a random noise-realization sequence to the noise-free OPD sequence. The measurement error $\sigma_{w}^{ij}$ on a baseline $ij$ is inversely proportional to the mean observed visibility $V^{ij}$ and the S/N $\rho^{ij}$ on that baseline, where
\begin{equation}
V^{ij}=\frac{2\sqrt{N^iN^j}}{N^i+N^j}\end{equation} and
\begin{equation}
\rho^{ij}=\frac{N^i+N^j}{\sqrt{N^i+N^j+4\cdot\textrm{RON}^2}}.
\end{equation}
Here, RON is the read-out noise per pixel and $N^i$ is the number of coherent photons arriving from telescope $i$ for a single phase measurement. The factor 4, on the other hand, is due to the phase sensing for GRAVITY being based on 4-pixel measurements (ABCD algorithm, see \citealt{1977JOSA...67...81S}). For simplicity, we only consider OPD estimation using phase-delay estimation. Phase-unwrapping techniques for resolving the $2\pi$ ambiguity in measured fringe phases, for example using a combination of phase-delay and group-delay algorithms (e.g.~\citealt{2011A&A...530A.121B}), are postponed to future work

In terms of the number of coherent photons $N$ arriving per exposure on one telescope, we have
\begin{equation}N^i=\frac13\cdot\frac15\,\mathcal T\,N,\end{equation}
where $\mathcal T$ is the total coherent throughput (AO-corrected atmospheric turbulence, delay-line system and GRAVITY). The factor $\frac13$ comes from the splitting of arriving photons on the three baselines corresponding to the telescope, while $\frac15$ refers to the dispersion of fringes on five spectroscopic channels (needed for the group-delay estimation, not considered in the current simulations). 

\begin{figure}[!t]
 \centering
 \includegraphics[width=0.49\textwidth,viewport=18 0 400 270, clip]{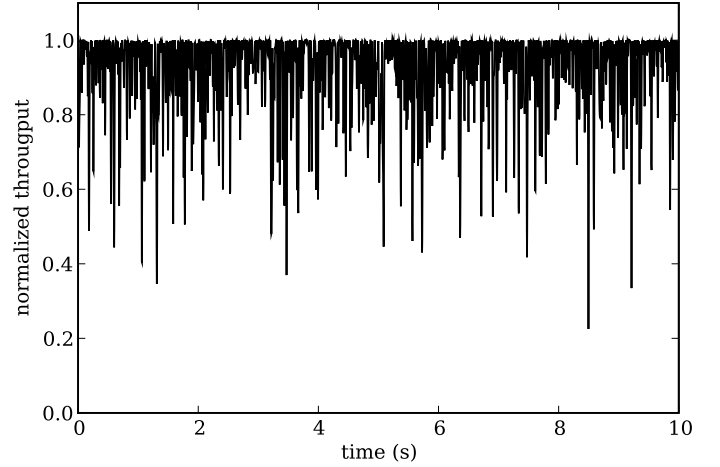}
 \caption{Ten-second extract of a normalized throughput sequence for one of the pistons.}\label{fig:dropout}
\end{figure}

The variability in the coherent throughput $\mathcal T$ is simulated based on a model for the propagating tip-tilt errors in the AO system, thus simulating improper fiber injection. Basically, we assume that
\begin{equation}\mathcal T=\mathcal T_0\,\exp\big(-\phi^2\big).\end{equation}
The quantity $\phi$ is the relative offset of proper fringe injection (the ratio of tip-tilt offset to the fiber mode field radius). On the basis of tip-tilt measurements at the VLTI, the temporal power spectrum for $\phi$ is derived to be
\begin{equation}S^\phi(f)\propto\left\{
\begin{array}{ll}
\log f&\textrm{for}\quad2\,\mathrm{Hz}<f<8\,\mathrm{Hz},\\
-\log f&\textrm{for}\quad8\,\mathrm{Hz}<f<50\,\mathrm{Hz},\,f\ne18.1\,\mathrm{Hz},\\
\delta(f)&\textrm{for}\quad f=18.1\,\mathrm{Hz}
\end{array}\right.\end{equation}
(P.~Gitton, private communication). The total rms tip-tilt deviation is scaled to 14.6\,mas, of which 5\,mas is present in the vibration component. A deeper discussion of this power spectrum would lead us too far for current purposes, and will be presented in future work (\citealt{2012...C}, in preparation). An extract of a simulated normalized throughput sequence is shown in Fig.~\ref{fig:dropout}.

Since phase delays estimated on the five spectroscopic channels can be combined, the estimator for the measurement error is derived to be\footnote{Note that, during operation, the number of photons per aperture can be directly estimated from the six ABCD measurements (ABCD photometry, see \citealt{2009A&A...498..601B}).}
\begin{equation}
\sigma_{w}^{ij}=\frac\lambda{2\pi}\sqrt{\frac25} \frac{\sqrt{N^i+N^j+4\cdot\textrm{RON}^2}}{2\sqrt{N^iN^j}}\label{eq:sigmawij}
\end{equation}
\citep{2008SPIE.7013E..1H}. The parameters in this and the above equations are specified in Table \ref{table:simulpar}, and are representative of the astrometric observation of the Galactic center, which is the principal science case of GRAVITY. At maximal throughput ($\mathcal T=\mathcal T_0$), we derive a coherent photon number per exposure of $N^i=20$ and a corresponding measurement error of $\sigma_w^{ij}=68\,$nm.

\begin{table}
 \centering
 \caption{Simulation parameters, representative for the Galactic-center observation of GRAVITY.}\label{table:simulpar}
 \begin{tabular}{lr}
  \hline\hline \textbf{Observation \& hardware}\\
  star magnitude K&10\\
  central wavelength $\lambda$&2.22\,$\upmu$m\\
  telescope diameter $D$&8.2\,m\\
  baseline $B$&80\,m\\
  sampling frequency $f$&300\,Hz\\
  tracking run&100\,s\\
  $\#$ spectral channels&5\\
  fringe-sensing algorithm&phase delay\\
  \hline \textbf{Disturbances \& noise}\\
  rms piston turbulence&10\,$\upmu$m\\
  wind speed $v$&15\,m\,s$^{-1}$\\
  $\#$ vibrations per baseline&8-10\\
  rms piston vibration&200-240\,nm\\
  max.~coh.~throughput $\mathcal T_0$&0.007\\
  rms tip-tilt deviation&14.6\,mas\\
  RON&6\,e$^{\mathrm-}$\\\hline
 \end{tabular}

\end{table}

\subsection{Turbulence-only simulations}
Before considering simulations with the complete set of expected OPD perturbations (turbulence and longitudinal vibrations), we perform a number of fringe-tracking runs including only the effect of turbulence (but with flux dropouts). The simulation process is started by generating a sequence of 2000 POL-data points. To keep the focus of our simulations on investigating the principle and performance of Kalman-filter fringe tracking, we chose to construct POL data in a simple way by adding measurement errors (Eq.~(\ref{eq:sigmawij})) to a turbulence OPD sequence. In complete and fully realistic simulations of the fringe tracking, the simulated POL data should be acquired in a frame-by-frame simulation with the phase-sensing algorithm. As mentioned before, this problem will be addressed in our future work (\citealt{2012...C}, in preparation).

Following the operation scheme defined in Sect.~\ref{sect:full}, we simulate 200 full fringe-tracking runs of 100\,s for both the modal and the OPD Kalman filters. To test our definition of the instantaneous Kalman gains when handling flux dropouts, we distinguish between simulations without modifying the gains (F, fixed) and with instantaneous gains (I).

Once the fit parameters are obtained and the Kalman matrices are generated, both recombination schemes can be tested. The fringe tracking is initiated with an empty state vector ($\hat{\mathbf x}_{0|-1}=\mathbf 0$). The convergence of the Kalman filter to the actual state of the system typically takes a few to a few tens of tracking points. We reject this transition region when calculating the residual OPD.

\begin{figure}[!t]
\centering
\includegraphics[width=0.49\textwidth, viewport=15 0 400 270, clip]{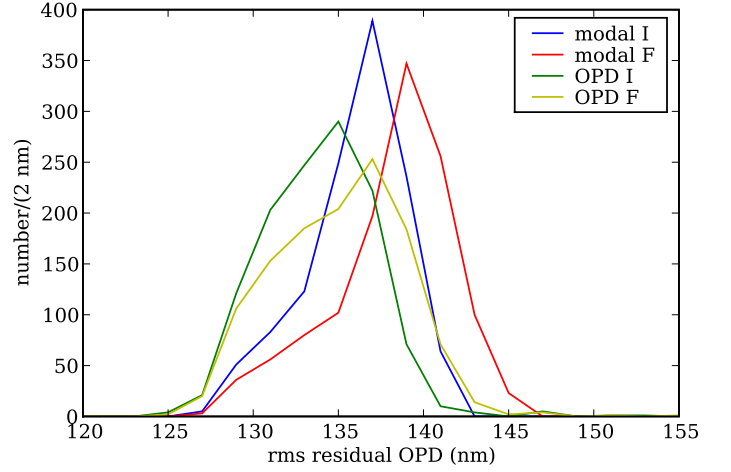}
\caption{Distribution of residual OPD values resulting from 200 simulations with turbulence as the only OPD disturbance. The plots show the number of residuals per block of 2\,nm. Both Kalman-filter schemes are considered, with instantaneous (I) and fixed (F) gains. }\label{fig:hist2000tur}
\end{figure}

The results of the turbulence-only simulations are shown in Fig.~\ref{fig:hist2000tur}. In this plot, all residual rms OPD values (6 per run) are grouped in blocks of 2\,nm, giving us a measure of the distribution of residual OPD values. We found that the performance of both control schemes (modal and OPD) in both gain settings (fixed and instantaneous) is similar. The residuals are distributed with some degree of asymmetry around a nominal residual value in the range of 125-145\,nm. The total tracking residuals of 100-s simulations are thus only about twice as large as the measurement error $\sigma_w^{ij}$ at maximum throughput for a single OPD observation ($68\,$nm, see previous section). This indicates that the tracking performance of the Kalman-filter schemes is good. For both filters (modal versus OPD), the gain modification slightly reduces the residual OPDs, by about 3-4\,nm (about 5-6\,$\%$ in energy).

\subsection{Full perturbation simulations}
We now turn to simulations based on the full power spectrum in Fig.~\ref{fig:profile}.  We put a maximum of 40 on the number of vibrations that can be identified during the mode identification, and 20 per baseline for the OPD identification. These numbers correspond to a maximum matrix size of $240\times240$ for $\mathsf\Sigma_\infty$.

\begin{figure}[!t]
\centering
\includegraphics[width=0.5\textwidth,viewport=0 0 400 340,clip]{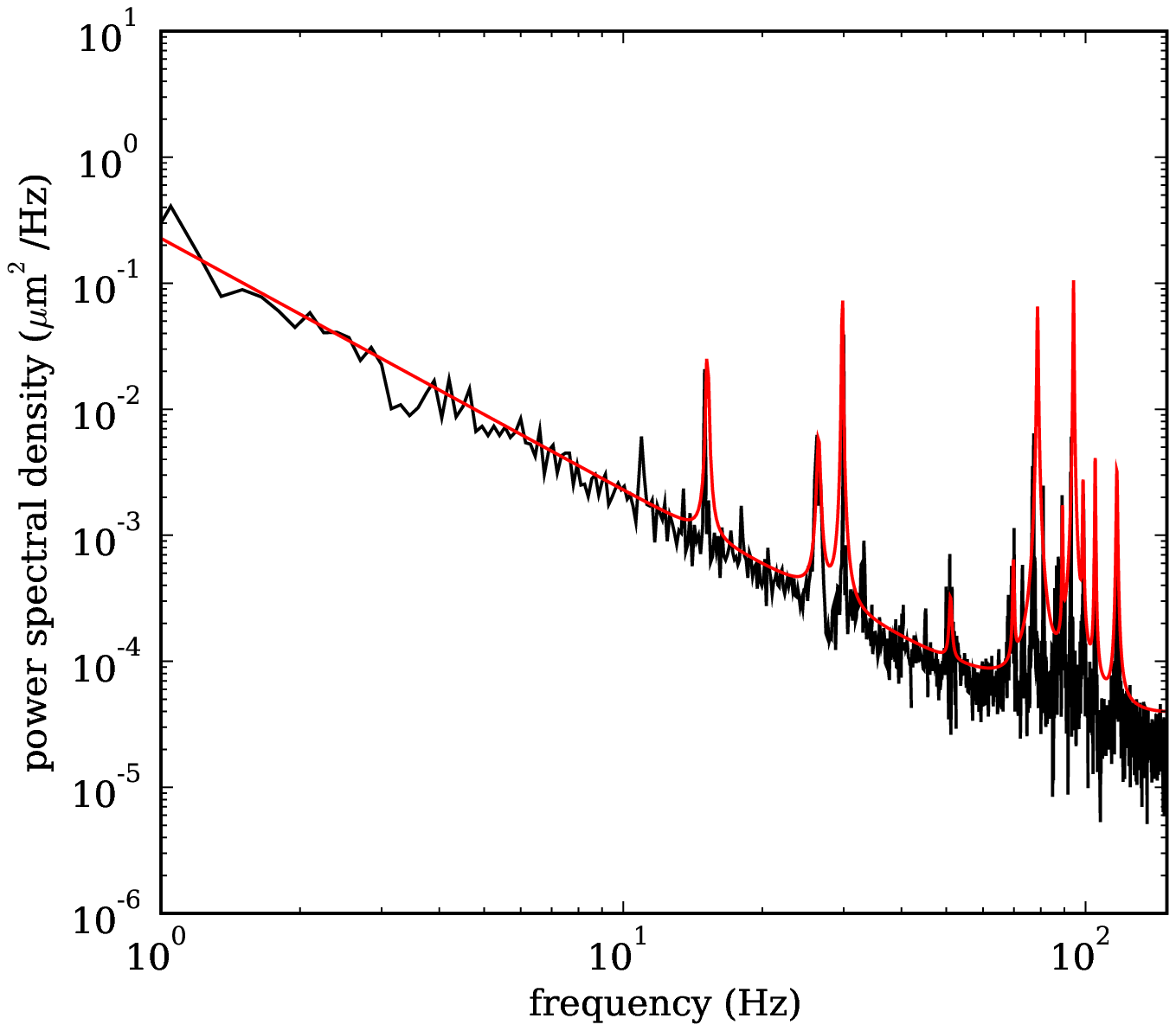}
\includegraphics[width=0.5\textwidth,viewport=0 0 400 340,clip]{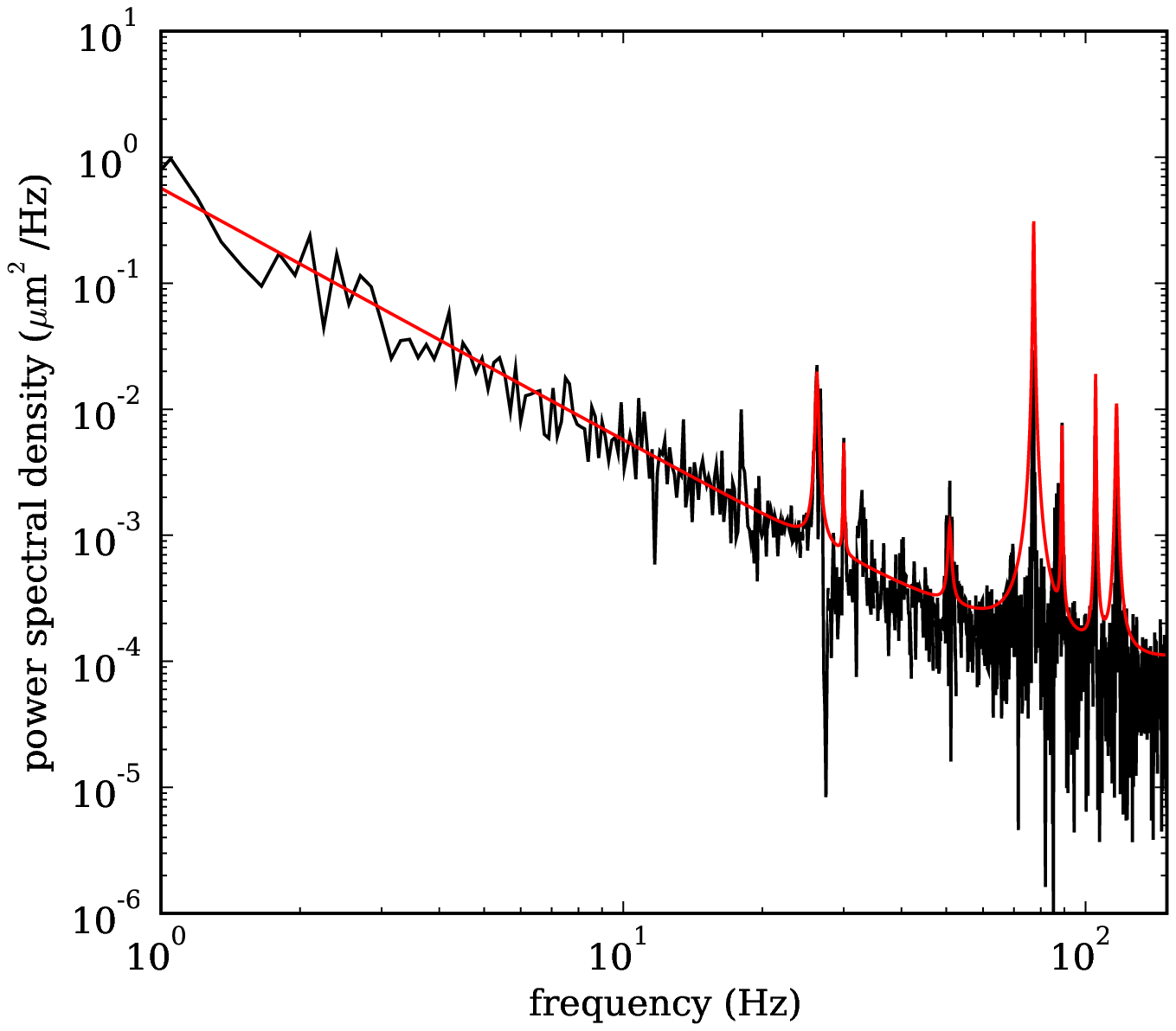}
\caption{Result of applying the spectral-fit method to an average mode periodogram (top) and one of the six OPD periodograms (bottom). The statistical spread in the periodogram points is clearly smaller than in the average periodogram, in other words the power spectrum is more accurately estimated.}\label{fig:perio_vib}
\end{figure}

As in the case of the turbulence-only simulations, we used 2000-point data sequences for the identification. The result of a typical spectral fit is shown in Fig.~\ref{fig:perio_vib}. By definition, the average mode spectrum contains all vibration components. This is clearly apparent in the upper plot, where the number of components to identify is about twice as high (a single baseline has the vibrations of only two telescopes). Another thing to note is that the statistical spread of the average modal periodogram is indeed lower than that of the OPD periodogram. 

\begin{figure}[!t]
\centering
\includegraphics[width=0.49\textwidth, viewport=15 0 400 270, clip]{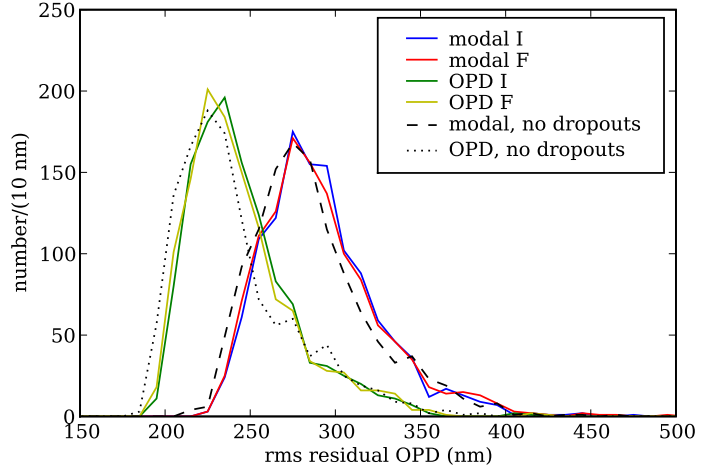}
\caption{Distribution of residual OPD values resulting from 200 simulations with full perturbation contribution. The plots show the number of residuals per block of 10\,nm. Both Kalman-filter schemes are considered, with instantaneous (I) and fixed (F) gains. In black dashed and dotted lines, we overplot the corresponding distributions for simulations without flux dropouts.}\label{fig:hist2000}
\end{figure}

A total of 200 full simulations of identification and $100$-s fringe-tracking run were performed. The corresponding distributions of OPD residuals are shown in Fig.~\ref{fig:hist2000}, organized into blocks of 10\,nm. The addition of vibration perturbations has introduced a considerable spread in the residual OPD values. In addition, unlike in the turbulence-only simulations, a clear difference in performance between the modal and the OPD scheme is found. The residual rms value for the modal scheme is about $290\,$nm, on average, where about $30\,\%$ of the measurements are above $300\,$nm. The OPD scheme, on the other hand, has an average of about $240\,$nm for the rms OPD value, and only $6\,\%$ of the residuals surpass $300\,$nm. When choosing between fixed or instantaneous gains, there is no apparent significant difference. We describe these observations in greater detail in Sect.~\ref{sect:discussion}.

Since the vibration perturbations considered are on the level of 200-240\,nm, the corresponding OPD vibration perturbations are roughly 300\,nm (e.g.~$220\sqrt2\,\mathrm{nm}=311\,$nm). Comparing the smallest residuals of turbulence-only simulations ($\sim125$\,nm) to the best residuals of the current simulations ($\sim200$\,nm), our simulations indicate that as much as $75\,\%$ of the vibration energy can be filtered out by our Kalman filters (we calculate that $(200^2-125^2)/311^2=25\,\%$ of the vibration energy is left).

\subsection{Discussion}\label{sect:discussion}
The above simulations give a first quantitative result for the Kalman-filter schemes developed in this work. We have illustrated the operation of the spectral fit on realistic perturbation profiles, and have found the performance one would expect from the modal and OPD control schemes.

When considering our definition of instantaneous gains, it is clear that no significant differences between fixed and instantaneous gains are seen, for the considered flux perturbation sequences. One conclusion that might be drawn is that the instantaneous gains do not improve the fringe-tracker performance. To verify that this would be the case, we performed simulations without the flux dropouts, for which we added the results to Fig.~\ref{fig:hist2000}. Surprisingly, although there is some apparent difference, the global pattern in the distribution of OPD residuals seems to be rather unaffected by the flux dropouts. At this point, it might be expected that the Kalman-filter based control scheme is not limited by the expected fluctuations in flux. Future comparison to other control strategies would be required to determine whether this property is exclusive to Kalman filtering. Although the instantaneous-gain definitions do not change the Kalman-filter performance for the considered flux perturbations, we assume that the gains are still useful. When poorer observing conditions prevail, causing longer flux dropouts for instance, the gain can still be expected to limit the risk of improper tracking. In this respect, using instantaneous gains might be very advantageous to the fringe tracking on the ATs, which lack a higher-order AO correcting system. Additionally, when lower fluxes are assumed for the observed object, the gain modification becomes useful.

The main observation of our simulations is that the OPD Kalman filter provides the best results. This is somewhat unexpected, since the modal scheme has the most reliably estimated power spectrum to model, and has the optimal number of variables to track. One of the main disadvantages of the modal scheme, however, is that all vibrations are fitted at once. As a result, the periodogram fit is more difficult: 
\begin{itemize}
 \item A higher fraction of the periodogram points are part of a vibration peak, thus fewer points are available to estimate the turbulence component. Since the latter component is the most important perturbation, a good estimation of its associated power spectrum is essential.
 \item Vibration peaks place a much stronger bias on the estimation of the turbulence component. The reason is that the statistics of an average mode periodogram are ``narrower'' (i.e.~have a smaller spread, see the discussion in Appendix \ref{appendix:details}), hence vibration peaks have a stronger influence on the turbulence estimation than in the case of an OPD-periodogram fit.
 \item The more peaks that are present in the periodogram, the higher the probability that the peaks will overlap. Owing to the insufficient resolution of the periodograms, overlapping peaks will be fitted as a single broader peak, which is more difficult to correct (more transient behavior).
\end{itemize}
All of these properties might lower the quality of the power-spectrum estimation for the modal scheme, hence lower the Kalman filter performance.

Another property that might have been incorporated to explain the performance differences between modal and OPD control is that flux dropouts have different implications for the schemes. We recall that a low level of flux at one telescope can significantly affect the quality of all modes, whereas three of the six OPDs are well-measured in the OPD scheme. However, since simulations with flux dropouts give similar results to simulations without, the performance difference cannot be attributed to flux dropouts. Temporally losing information about all mode observables thus does not significantly influence the tracking performance of the Kalman filter. In contrast, for non-predictive control algorithms applied to the modes, losing observables can be expected to be a critical limitation.

When considering the actual shape of the distributions in Fig.~\ref{fig:hist2000}, we find a quite large spread in the residual OPD values. Not taking into account local statistical jitter, the distributions have a clear global maximum with some ``leaking'' to higher residuals ($\sim10$-$15\,\%$). We ascribe these outliers to improper estimation of the power spectra of the perturbation components. Improvements to the identification method could indeed be made to make it more robust (e.g.~optimal grid construction, reconsidering the number of identification points). 

As a final note, it cannot be ruled out that the performance difference is related to the conceptual difference between the Kalman filters. For a vibration acting on one telescope, the three baselines that are not associated remain unaffected from this vibration. In the modal scheme, however, the vibration needs to be filtered in all modes. It might be this property of indirectly affecting the other telescopes (via the modal filtering) that causes the lower average performance of the modal scheme. More tests are needed to verify this reasoning.

\subsection{Performance with respect to GRAVITY} 
The main conclusion of these simulations is that the average performance of the Kalman-filter control schemes is compatible with the performance required for GRAVITY, specified at $300\,$nm \citep{2010SPIE.7734E..28G}. For the OPD scheme, at least $90\,\%$ of the performed tracks should attain this stability level, and a nominal performance of $\lambda/10$ is within reach. This makes the latter scheme clearly the most suitable of the two.

The most important working point to reach this stability level for our Kalman-filter schemes is optimizing the operation of the identification method. In this respect, a decent implementation should limit both the risk of low-quality identification and the calculation time. For the latter point, an optimal adaptation of the parameter grid to the typical prevailing disturbances at the VLTI should be defined (\citealt{2012...C}, in preparation).

Compared to the simulations done in \citet{2010SPIE.7734E..59C}, based on a classical integrator controller, our simulations show that the nominal Kalman-filter performance might be compatible or better. In any case, we should immediately nuance this comparison, since the assumed parameters and perturbation profiles are rather different than the ones used here. For example, the simulated turbulence profiles used here exclude the unobserved $f^{-17/3}$ part and a higher and more realistic vibration contribution is assumed. Our future work will present a firmer base for comparing Kalman-filter fringe tracking to classical control, something that is beyond the reach of the current work.

\section{Summary}
In the framework of new-generation interferometric instruments, we have designed and analyzed a Kalman control scheme for fringe tracking. The starting point of this work was the work done in AO control, in particular a control scheme proposed by \citet{2004JOSAA..21.1261L} based on Kalman filtering, which is an optimal data processing algorithm. Apart from the usual correction for turbulence disturbance, a Kalman filter allows us to correct for vibration disturbances, present in VLTI observations under the form of longitudinal vibrations.

We started by formulating a control scheme for two-telescope fringe tracking, based on a parametric model for the fringe-tracking disturbances. To estimate the involved parameters, we included an identification method based on the characterization of perturbation power spectra \citep{2010JOSAA..27.122M}. As a direct application to multi-telescope fringe tracking, we extended the fringe-tracking formalism to four-telescope control schemes. We note that our approach can easily be generalized to $n$-telescope fringe tracking.

Two Kalman-filter control schemes are considered in the four-telescope case: a modal-based scheme, where three modes (linear combinations of pistons) are tracked in a single three-component Kalman filter, and an OPD-based scheme, where six OPDs are tracked using six Kalman filters. The advantage of the former is that only one spectral fit needs to be performed: the three modal periodograms can be averaged. To ensure that the control schemes are unaffected by low S/N baselines and low fluxes at the level of the telescopes (flux dropouts), we included a weighting step and a modification to the gain. 

Simulations performed in the context of the four-telescope beam combiner VLTI/GRAVITY allowed to test the two control schemes and the concept of instantaneous gains. Two important observations that follow are:
\begin{enumerate}
 \item Despite being operationally heavier, the OPD control scheme provides the best results. We (partly) attribute the lower performance of the modal control to the more difficult identification (all vibration parameters need to be extracted at once). There might also be a conceptual performance limitation related to Kalman filtering of the modes, in that all perturbations need to be corrected in all modes.
 \item For the considered flux perturbations, the instantaneous gains do not significantly improve the fringe-tracking performance. 
\end{enumerate}
When considering the latter point, comparison to simulations without the flux dropouts have indicated that Kalman-filter control seems to be rather unaffected by the considered flux perturbations. Yet, the instantaneous gain might prove useful when worse observing conditions prevail (e.g.~fainter tracking source, longer flux dropouts).

As far as the science case of VLTI/GRAVITY is concerned, both schemes show nominal performances that meet the specified performances. The results open perspectives for attaining even higher quality residuals ($\lambda/10$ performance). The most important working point for the future is optimizing the identification method.

\begin{acknowledgements}
We would like to thank Pierre Fedou for useful interactions on the design of VLTI/GRAVITY. We are grateful to Frank Eisenhauer, the editor (Thierry Forveille) and the anonymous referee for comments and suggestions that helped improving this manuscript.
\end{acknowledgements}

\bibliographystyle{aa}
\bibliography{references.bib}

\appendix

\section{Derivation of the modal state-space model}\label{appendix:modalKalman}
The derivation below is based on the state-space model in Sect.~\ref{sect:Kalman2}. We use a natural intermediate state-space model to develop the modal model.

\subsection{Canonical state-space model} Before going to the complete set of four telescopes, we recast the problem of a two-telescope interferometer (Sect.~\ref{sect:Kalman2}) in a more useful form. The result allows a direct generalization to an $n$-telescope interferometer.

\subsubsection{Two-telescope interferometer in a piston scheme} 
In two dimensions, piston disturbances and associated OPD measurements can be written like $\mathbf P =(P^0\,\,P^1)^\top$ and (trivially) $\mathbf{OPD}=(\mathrm{OPD}^{01})$. Finally, the matrix $\mathsf M$ that governs the conversion $\mathbf P\rightarrow \mathbf{OPD}$ is
\begin{equation}\mathsf M=\big(-1\,\,\,\,\,1\big).\end{equation}
We define a new state-space model in terms of fundamental piston disturbances $\mathbf P$ (rather than the phase disturbances $\varphi$ in Sect.~\ref{sect:Kalman2}). For instance, a vibration on the level of piston $P^k$ ($k=0,1$) can be written as
\begin{equation}P^{k,\mathrm{vib}}_{n+1}=a^{k,\mathrm{vib}}_{1}P^{k,\mathrm{vib}}_{n}+ a^{k,\mathrm{vib}}_{2}P^{k,\mathrm{vib}}_{n-1}+v^{k,\mathrm{vib}}_{n},\end{equation}
where we note that we simply include one extra index, $k$, to refer to the considered piston. In this state-space model, new state vectors $\mathbf x$ are defined as
\begin{equation}
\mathbf x_n=\big(\mathbf x^0_n\,\,\,\mathbf
x^1_n\big)^\top,\label{eq:example2}
\end{equation}
where, taking e.g.~two vibrations on piston 0 and one on piston 1,
\begin{eqnarray*}
\mathbf
x^0_n&=&\big(P^{0,\mathrm{tur}}_{n}\,\,\,P^{0,\mathrm{tur}}_{n-1}
\,\,\,\,P^{0,\mathrm{vib\,1}}_{n}\,\,\,P^{0,\mathrm{vib\,1}}_{n-1}\,\,\,P^{0,\mathrm{vib\,2}}_{n}\,\,\,P^{0,\mathrm{vib\,2}}_{n-1}\big)^\top,\\
\mathbf
x^1_n&=&\big(P^{1,\mathrm{tur}}_{n}\,\,\,\,P^{1,\mathrm{tur}}_{n-1}\,\,\,\,P^{1,\mathrm{vib\,1}}_{n}\,\,\,P^{1,\mathrm{vib\,1}}_{n-1}\big)^\top.
\end{eqnarray*}
The full equation of state can then be written in block form as
\begin{equation}
\underbrace{\begin{pmatrix} \mathbf x^0_{n+1}\\\mathbf x^1_{n+1}
\end{pmatrix}}_{\mathbf x_{n+1}}=\underbrace{\begin{pmatrix}
\mathsf A^0&0\\0&\mathsf A^1
\end{pmatrix}}_{\mathsf A}\underbrace{\begin{pmatrix}
\mathbf x^0_{n}\\\mathbf x^1_{n}
\end{pmatrix}}_{\mathbf x_n}+\underbrace{\begin{pmatrix}
\mathbf v^0_n\\\mathbf v^1_n
\end{pmatrix}}_{\mathbf v_n},\label{eq:eos2}
\end{equation}
where block diagonal matrices $\mathsf A^i$ are defined exactly as before (for each fundamental piston). We assume that the system noise $\mathbf v$ is composed of independent white Gaussian noise components, hence the covariance matrix $\mathsf\Sigma_v$ is still diagonal.

Given that we consider the same two-telescope interferometer as before, the observable is still the same: the residual OPD $y^{01}_n$ (note that we give the index ``01'' to refer to the baseline on which we work). The new state vector requires an updated definition of the system matrix $\mathsf C$, which we denote $\overline{\mathsf C}$. An appropriate definition is
\begin{equation}
\overline{\mathsf C}=(-\mathsf C^0\,\,\,\,\,\mathsf
C^1)=(-1\,\,\,\,\,1)\begin{pmatrix}\mathsf C^0&0\\0&\mathsf
C^1\end{pmatrix}=\mathsf M\underbrace{\begin{pmatrix}\mathsf
C^0&0\\0&\mathsf C^1\end{pmatrix}}_{\mathsf C}\label{eq:C2},
\end{equation}
where $\mathsf C^0$ and $\mathsf C^1$ are row matrices defined as before. For example, the corresponding $\mathsf C^0$ and $\mathsf C^1$ to Eq.~(\ref{eq:example2}) are
\begin{equation}\mathsf C^0=(0\,\,\,1\,\,\,0\,\,\,1\,\,\,0\,\,\,1)\quad\textrm{ and }\quad
\mathsf C^1=(0\,\,\,1\,\,\,0\,\,\,1).\end{equation} 
In accordance with the definition of the matrix $\mathsf A$ in Eq.~(\ref{eq:eos2}), we have defined $\mathsf C$ in Eq.~(\ref{eq:C2}) as the block-diagonal matrix of matrices $\mathsf C^i$. Considering the piston control, we wish to calculate the commands to be applied directly to the pistons. Command vectors are hence of length two (we consider two actuators). The full measurement equation in its new form is then
\begin{equation}
y^{01}_n=\mathsf M\,\mathsf C\mathbf x_n-\mathsf M\,\mathbf
u_{n-2}+w^{01}_{n}.\label{eq:oe2}
\end{equation}

\subsubsection{Four-telescope interferometer in piston scheme} 
The equation of state in Eq.~(\ref{eq:eos2}) in now trivially generalized to the four-double equivalent
\begin{equation}
\underbrace{\begin{pmatrix} \mathbf x^0_{n+1}\\\mathbf
x^1_{n+1}\\\mathbf x^2_{n+1}\\\mathbf x^3_{n+1}
\end{pmatrix}}_{\mathbf x_{n+1}}=\underbrace{\begin{pmatrix}
\mathsf A^0&0&0&0\\0&\mathsf A^1&0&0\\0&0&\mathsf
A^2&0\\0&0&0&\mathsf A^3
\end{pmatrix}}_{\mathsf A}\underbrace{\begin{pmatrix}
\mathbf x^0_{n}\\\mathbf x^1_{n}\\\mathbf x^2_{n}\\\mathbf x^3_{n}
\end{pmatrix}}_{\mathsf x_n}+\underbrace{\begin{pmatrix}
\mathbf v^0_n\\\mathbf v^1_n\\\mathbf v^2_n\\\mathbf v^3_n
\end{pmatrix}}_{\mathbf v_n}.\label{eq:eos4}
\end{equation}
We still assume that the covariance matrix $\mathsf\Sigma_v$ is diagonal, i.e.~the vectors $\mathbf v^k_n$ ($k=0,1,2,3$) have independent white Gaussian noise components. The fact that fringes are tracked on six baselines leads to the use of residual OPD vectors $\mathbf y$ and measurement noise vectors $\mathbf w$, where
\begin{equation} \mathbf
y=\big(y^{01}\,\,\,y^{02}\,\,\,y^{03}\,\,\,y^{12}\,\,\,y^{13}\,\,\,y^{23}\big)^\top,
\end{equation}
and analogously for $\mathbf w$. The system matrix $\mathsf C$ and the vectors $\mathbf u_n$ also have their four-dimensional equivalent, and the full measurement equation finally reads
\begin{equation}
\mathbf y_{n}=\mathsf M\,\mathsf C\,\mathbf x_n-\,\mathsf M\,\mathbf
u_{n-2}+\mathbf w_{n}.\label{eq:oe4}
\end{equation}
As we did with $\mathsf\Sigma_v$, the measurement error covariance matrix $\mathsf\Sigma_w$ is assumed to be diagonal.

\subsubsection{Problems with the state-space model} The combination of Eq.~(\ref{eq:eos4}) and Eq.~(\ref{eq:oe4}) forms a very natural full state-space model of a four-telescope six-baseline fringe tracking system. On the one hand, it describes in a natural way the disturbances on the level of the four apertures, under the reasonable assumption that these are decoupled. On the other hand, the measurement equation describes how the six measured residual OPDs are the result of the contributions of the four fundamental pistons, and of the correction that is applied by the four actuator systems (piezos + delay lines). Owing to the very natural description given by this model, we can refer to it as the \emph{canonical} state-space\ model. It seems that we have found an appropriate input model to build a Kalman filter.

However, there is a serious difficulty of the canonical model: it is impossible to identify its parameters. In Sect.~\ref{sect:identification}, the parameter identification method that is introduced is based on applying a spectral fit to sequences of (pseudo-)open loop measurements. Here, however, the matrix $\mathsf M$ couples all information about the evolution of pistons, in a non-trivial way. In essence, the problem is that absolute piston information cannot be unambiguously recovered from observations. Difficulties with the system identification for a piston-based state-space model could hence have been expected. However, the state-space model defined above now forms a perfect starting point for the construction of a modal-based state-space model.

\subsection{Modal state-space model}\label{sect:modalstatespacemodel}
We now consider how the canonical state-space model can be put into a modal form, in the spirit of what was done in Sect.~\ref{sect:modalbasedcontrol}. This direct derivation of a modal state-space model is more reliable than simply applying one Kalman-filter controller per mode. In particular, although the modes are orthogonally defined, they have coupling terms that are related to there being linear combinations of pistons/OPDs. The power of the Kalman filter is that it is constructed to deal with vector quantities and covariance matrices which describe coupling between different terms.

\paragraph{Modal equation of state.} For the new equation of state, we introduce a piece of new notation. We use the tilde sign ($\sim$) to denote extended vector and matrix quantities with respect to the previously introduced versions. First we introduce the extended state-space vectors $\widetilde{\mathbf x}_n$ as
\begin{equation}
\widetilde{\mathbf x}_n=(\widetilde{\mathbf
x}^0_n\,\,\,\,\widetilde{\mathbf x}^1_n\,\,\,\,\widetilde{\mathbf
x}^2_n\,\,\,\,\widetilde{\mathbf x}^3_n)^\top,\label{eq:xtilde}
\end{equation}
where e.g.
\begin{equation}
\widetilde{\mathbf x}^2_n=\big(\mathbf 0^0\,\,\,\,\,\mathbf
0^1\,\,\,\,\mathbf x^2_n\,\,\,\,\mathbf 0^3\big)^\top, \qquad (\mathbf 0^j=0\,\mathbf x^j_n\,\textrm{ for }\, j=0,1,3).
\end{equation}
In words, the components $\widetilde{\mathbf x}^i_n$ consist of four zero-columns of which the $k$-th zero-column has the same length as the vector $\mathbf x^k_n$ ($k=0,1,2,3$), but the $i$-th zero-column is replaced by $\mathbf x^i_n$ itself. It follows that, by definition, all $\widetilde{\mathbf x}^i_n$ have the same length (unlike the vectors $\mathbf x^i_n$), and we define
\begin{equation}r\equiv\textrm{length of vectors }\widetilde{\mathbf x}^i_n.\end{equation}
In exactly the same way as in Eq.~(\ref{eq:xtilde}), we define the vectors $\widetilde{\mathbf v}_n$ as extensions to $\mathbf v_n$, and the corresponding covariance matrix is written as $\mathsf\Sigma_{\widetilde v}$. Finally, we define the matrix $\widetilde{\mathsf A}$ as
\begin{equation}
\widetilde{\mathsf A}=\begin{pmatrix} \mathsf A&\mathsf 0&\mathsf
0&\mathsf 0\\\mathsf 0&\mathsf A&\mathsf 0&\mathsf 0\\\mathsf
0&\mathsf 0&\mathsf A&\mathsf 0\\\mathsf 0&\mathsf 0&\mathsf
0&\mathsf A
\end{pmatrix},
\end{equation}
i.e.~a block diagonal matrix with the square system matrix $\mathsf A$ on the diagonal. A completely equivalent equation of state to the one in Eq.~(\ref{eq:eos4}) is then given by
\begin{equation}
\widetilde{\mathbf x}_{n+1} =\widetilde{\mathsf A}\,
\widetilde{\mathbf x}_n+ \widetilde{\mathbf v}_n.\label{eq:eostilde}
\end{equation}
To pass to a modal equation of state, we follow an approach similar to that used in the Q-mode definition $\mathbf Q=\mathsf V^\top\,\mathbf P$. We define a (still unitary) block-matrix version of the matrix $\mathsf V$
\begin{equation}
\mathsf V_\square=\frac12\begin{pmatrix} -\mathsf 1&-\mathsf 1&-\mathsf 1&\mathsf 1\\
-\mathsf 1&\mathsf 1&\mathsf 1&\mathsf 1\\
\mathsf 1&-\mathsf 1&\mathsf 1&\mathsf 1\\
\mathsf 1&\mathsf 1&-\mathsf 1&\mathsf 1
\end{pmatrix},\end{equation}
where ``$\mathsf 1$'' denotes a unit $r\times r$ matrix. The modal state vector is then defined as
\begin{equation}
\mathbf x_{Q,n}\equiv\mathsf V_\square^\top\,\widetilde{\mathbf
x}_n.
\end{equation}
Multiplying Eq.~(\ref{eq:eostilde}) by $\mathsf V_\square^\top$ and using the equality $\widetilde{\mathsf A}=\mathsf V_\square^\top\widetilde{\mathsf A} \mathsf V_\square$, the modal equation of state finally reads
\begin{equation}
\mathbf x_{Q,n+1} =\widetilde{\mathsf A}\,\mathbf x_{Q,n}+\mathbf
v_{Q,n},\label{eq:eosmode}
\end{equation}
where we have defined the mode system noise
\begin{equation}
\mathbf v_{Q,n}\equiv\mathsf V_\square^\top\,\widetilde{\mathbf v}_n.
\end{equation}
The covariance matrix of $\mathbf v_Q$ is $\mathsf\Sigma_{Q,v}=\mathsf V_\square^\top\,\mathsf\Sigma_{\widetilde v}\,\mathsf V_\square$. Using some numerical examples, it can be shown that $\mathsf\Sigma_{Q,v}$ has a quadruple diagonal structure, i.e.~four equal diagonal blocks, but also possesses off-diagonal terms.

\paragraph{Modal observation equation.} We multiply the observation equation by $\mathsf H$ (Eq.~(\ref{eq:Hmatrix})), which leads to
\begin{equation}
\mathbf y_{Q,n}\equiv\mathsf H\,\mathbf y_n=\mathsf R\,\mathsf
V^\top\mathsf C\mathbf x_n-\mathsf R\,\mathsf V^\top\mathbf
u_{n-2}+\mathsf H\,\mathbf w_{n},\label{eq:oemodified}\end{equation}
where $\mathsf R=\mathrm{diag}\big(1,1,1,0\big)$. We note that we used the property $\mathsf H\,\mathsf M=\mathsf R\,\mathsf V^\top$. We verify that this equation indeed defines the last component of $\mathbf y_{Q,n}$ to be zero, in accordance with our convention that $Q^3\equiv0$. The defined transformation gives the observing equation in terms of the residual Q-mode observables $\mathbf y_Q$. Next, we need to pass from the piston state vector $\mathbf x_n$ to the equivalent $\mathbf x_{Q,n}$ for the modes. For this, we define an extended version of the matrix $\mathsf C$ as\footnote{The horizontal bar indicates that $\mathsf C_-$ is a row vector built of row vectors, and allows us to distinguish from the previously introduced $\mathsf C$ matrix (Eq.~(\ref{eq:C2})). In the principal text, we drop this bar, for notational
simplicity.}
\begin{equation}\widetilde{\mathsf C}=\begin{pmatrix} \mathsf
C_-&\mathsf 0&\mathsf 0&\mathsf 0\\\mathsf 0&\mathsf C_-&\mathsf
0&\mathsf 0\\\mathsf 0&\mathsf 0&\mathsf C_-&\mathsf 0\\\mathsf
0&\mathsf 0&\mathsf 0&\mathsf
0\end{pmatrix},\qquad\textrm{where}\qquad \mathsf C_-=(\mathsf
C^0\,\,\,\,\mathsf C^1\,\,\,\,\mathsf C^2\,\,\,\,\mathsf C^3).\end{equation} It takes some pen and paper work to show that one can rewrite the first term in the right hand side of Eq.~(\ref{eq:oemodified}) as
\begin{equation}\mathsf R\,\mathsf V^\top\,\mathsf C\,\mathbf x_n=\widetilde{\mathsf C}\,\mathbf x_{Q,n}.\end{equation}
Two final definitions are the modal command and the mode measurement noise
\begin{equation}
\mathbf u_{Q,n}\equiv \mathsf R\,\mathsf V^\top\,\mathbf
u_n\qquad\textrm{and}\qquad\mathbf w_{Q,n}\equiv\mathsf H\,\mathbf
w_{n}\end{equation} 
(for both, the last component is 0). The full modal equation is finally given by
\begin{equation}
\mathbf y_{Q,n}=\widetilde{\mathsf C}\,\mathbf x_{Q,n}-\mathbf
u_{Q,n-2}+\mathbf w_{Q,n}.\label{eq:oemode}
\end{equation}
We note that the covariance matrix of $\mathbf w_{Q}$ is $\mathsf\Sigma_{Q,w}=\mathsf H\,\mathsf\Sigma_w\,\mathsf H^\top$.

\section{Details of four-telescope identification}\label{appendix:details}
We denote by $P^y$ the periodogram of a POL-data sequence on a single baseline, i.e.~as in Eq.~(\ref{eq:ypol}). In addition, we define $S$ as the power spectrum associated with the periodogram $P^y$. Under the assumption of Gaussian white noise excitation, it can be shown that, for fixed frequencies $f$, the distribution of $P^y(f)$ asymptotically approaches an exponential distribution of expectation value $S(f)$. In other words, for long identification sequences, the likelihood function $\mathrm L$ of $P^y$ is close to
\begin{equation}
\mathrm
L(P^y,\boldsymbol\lambda)=\prod_f\frac1{S(f)}\exp\left(-\frac{P^y(f)}{S(f)}\right),\label{eq:likelihood}
\end{equation}
where $\boldsymbol\lambda$ denotes all model parameters of $S$.

Averaging three exponential distributions leads to a Gamma-Erlang distribution of order three. Hence, denoting by $\overline{P^y}$ the averaged periodogram, the periodogram likelihood function in Eq.~(\ref{eq:likelihood}) is modified into
\begin{equation}
\mathrm L(\overline{P^y},\boldsymbol\lambda)=\prod_{f}\frac12\left(\frac3{S(f)}\right)^3\big(\overline
{P^y}(f)\big)^2\exp\left(-\frac{3\overline
{P^y}(f)}{S(f)}\right).\label{eq:gammalikelihood}
\end{equation}
At a fixed frequency $f$, the expectation value of the averaged periodogram remains the same, whereas the standard deviation is reduced by a factor $\sqrt3$. The smaller standard deviation is the major advantage of periodogram averaging: the power spectrum is more accurately estimated, and even low vibration peaks become significant. The spectral-fit method then requires the following three modifications:
\begin{enumerate}
\item The maximum-likelihood criterion needs to be adapted to the likelihood function in  Eq.~(\ref{eq:gammalikelihood}).
\item The threshold for vibration-peak detection (i.e.~peak-picking criterion) can be lowered, since lower peaks become significantly detectable. More concretely, we choose the detection criterion $P^y(f)>4\,S(f)$ to detect vibration peaks (probability of $0.05\,\%$ per point for false detection), rather than the detection criterion $P^y(f)>7\,S(f)$ for non-averaged periodograms (probability of $0.1\,\%$ per point for false detection).
\item A more subtle point is the characterization of the measurement-noise covariance matrix $\mathsf\Sigma_{Q,w}$ (see Eq.~(\ref{eq:SigmaQw})). In the standard MPFK method, applicable to single-baseline data, the measurement error is determined from the horizontal tail of the POL-data periodogram. In modal periodograms, the information of the different baselines is mixed. As a solution, we can first extract the measurement errors of the 6 OPD periodograms. Alternatively, we can simply combine the measurement errors estimated by the phase sensor (see Eq.~(\ref{eq:sigmawij})) into a global measurement error (one for each baseline). We note that the measurement noise of the average modal periodogram is calculated by summing the diagonal of $\mathsf\Sigma_{Q,w}$ and dividing by three (i.e.~averaging over the three modes).
\end{enumerate}

\section{Instantaneous-gain definition}\label{appendix:instgain}
The calculation of an arbitrary Kalman gain matrix $\mathsf G_\infty$ (e.g.~as in Eq.~(\ref{eq:gain})) is of the form
\begin{equation}
\mathsf G_\infty=\mathsf\Sigma_\infty\mathsf C^\top(\mathsf
C\mathsf\Sigma_\infty\mathsf
C^\top+\mathsf\Sigma_{w})^{-1},\label{eq:regain}
\end{equation}
where $\mathsf\Sigma_\infty$ is calculated by solving a Riccati equation. It is clear that the gain depends on a measurement covariance $\mathsf\Sigma_{w}$. Every Kalman step should ideally involve the calculation of a new gain, to compensate for the variable measurement noise (flux dropouts). As this would be a heavy computational burden, we propose adding an artificial term to the gain. 

It is clear from Eq.~(\ref{eq:regain}) that the gain is inversely related to the $\mathsf\Sigma_{w}$ (neglecting that it is also used to calculate $\mathsf\Sigma_\infty$). It then makes sense to define an instantaneous gain $\mathsf G_{\infty,n}$ as a quantity of the form
\begin{equation}
\mathsf G_{\infty,n}\equiv\frac{f\big(\mathsf\Sigma_{w}\big)}
{f\big(\mathsf\Sigma_{w,n}\big)}\,\mathsf G_\infty,\label{eq:Ginftyn}
\end{equation}
where $f$ is a real function. The matrix $\mathsf\Sigma_{w,n}$ is defined as the instantaneous measurement covariance at time step $n$. For flux of the usual level of noise, the definition of $\mathsf G_{\infty,n}$ implies that we are tracking using the optimal gain $\mathsf G_\infty$ (the multiplicative term is 1). Whenever the instantaneous measurement noise increases, the multiplicative factor must ensure that the gain is decreased. In the extreme case when the measurements are very noisy and hence unusable, $f$ can be chosen such that the gain approaches 0. In that case, the Kalman filter is based completely on the deterministic part of the evolution.

For the modal and the OPD Kalman filter, we take the instantaneous-gain definitions as in Sect.~\ref{sect:fluxdropouts}. It is important to understand the difference between the modal and the OPD scheme that led to Eqs.~(\ref{eq:Ginftynmode}) and (\ref{eq:GinftyijnOPD}). In the modal scheme, each mode estimate is limited by the noise on the worst OPD triplet (by triplet, we mean three baselines that are connected to the same telescope). This is because these modes are defined as mixes of OPDs associated with all telescopes. A first order multiplicative correction of the gain thus preferably contains a measure for all error terms, hence we consider the trace of the covariance matrix, neglecting the off-diagonal terms. For the OPD scheme, however, each Kalman filter tracks a variable in an independent way. The gain of the different Kalman filters can therefore be adapted purely as a function of the noise on the corresponding variable. Again neglecting the coupling terms, the gain associated with a baseline $ij$ is modified by the corresponding term in the error covariance matrix. In this way, we avoid that the flux problems for one telescope propagating to the tracking on baselines not associated with that telescope.

We remark that the discussion that led us to Eqs.~(\ref{eq:Ginftynmode}) and (\ref{eq:GinftyijnOPD}) is probably not the end of the story. It is possible that more elegant ``instantaneous gain'' definitions exist. Yet, these definitions would likely have a similar form to our choice.

\end{document}